\documentclass[useAMS,usenatbib]{mn2e}
\usepackage{graphicx}
\def\eg{{\it e.g.}}
\def\etal{{\it et al.}}
\def\etc{{\it etc.}}
\def\ie{{\it i.e.}}

\def\DF{{\sc DF}}
\def\PK{{\sc PKDGRAV}}

\def\bL{\;\pmb{\mit L}}
\def\pmb#1{\setbox0=\hbox{$#1$}%
  \kern-0.25em\copy0\kern-\wd0
  \kern.05em\copy0\kern-\wd0
  \kern-0.025em\raise.0433em\box0}

\title[Stochasticity in $N$-body Discs]{Stochasticity in
$N$-body Simulations of Disc Galaxies}
\author[J. A. Sellwood and V.
P. Debattista]{J. A. Sellwood$^{1}$\thanks{E-mail:
sellwood@physics.rutgers.edu} and Victor
P. Debattista$^{2}$\thanks{E-mail: vpdebattista@uclan.ac.uk}\\
$^{1}$Rutgers University, Department of Physics \& Astronomy, 136
Frelinghuysen Road, Piscataway, NJ 08854-8019, USA\\ 
$^{2}$Jeremiah Horrocks Institute for Astrophysics and Supercomputing,
University of Central Lancashire, Preston, PR1 2HE, UK}
\begin{document}


\pagerange{\pageref{firstpage}--\pageref{lastpage}} \pubyear{2009}

\maketitle

\label{firstpage}

\begin{abstract}
We demonstrate that the chaotic nature of $N$-body systems can lead to
macroscopic variations in the evolution of collisionless simulations
containing rotationally supported discs.  The unavoidable
stochasticity that afflicts all simulations generally causes mild
differences between the evolution of similar models but, in order to
illustrate that this is not always true, we present a case that shows
extreme bimodal divergence.  The divergent behaviour occurs in two
different types of code and is independent of all numerical
parameters.  We identify and give explicit illustrations of several
sources of stochasticity, and also show that macroscopic variations in
the evolution can originate from differences at the round-off error
level.  We obtain somewhat more consistent results from simulations in
which the halo is set up with great care compared with those started
from more approximate equilibria, but we have been unable to eliminate
diverging behaviour entirely because the main sources of stochasticity
are intrinsic to the disc.  We show that the divergence is only
temporary and that halo friction is merely delayed, for a substantial
time in some cases.  We argue that the delays are unlikely to arise in
real galaxies, and that our results do not affect dynamical friction
constraints on halo density.  Stochastic variations in the evolution
are inevitable in all simulations of disc-halo systems, irrespective
of how they were created, although their effect is generally far less
extreme than we find here.  The possibility of divergent behaviour
complicates comparison of results from different workers.
\end{abstract}


\section{Introduction}
\citet{Mill64} pointed out that all gravitational $N$-body systems are
chaotic, in the sense that the trajectories of all particles in two
systems that differ initially by a small shift in the starting
position or velocity of even a single particle will diverge
exponentially over time.  Thus, two simulations started from the same
initial conditions will follow identical evolutionary paths only if
the arithmetic operations are performed with the same precision and in
the same order, so that round off error is identical.  These
statements are true for every code, irrespective of the algorithm used
for the computations, and no matter how many particles are employed.
In particular, a simulation can never be reproduced exactly when run
with a different code.

Microscopic chaos is unimportant for many applications because the
different evolutionary paths of almost identical simulations lead to
similar macroscopic properties such as mass profiles, overall shape,
\etc, which therefore constitute firm results.  \citet[][hereafter
BT08, p.~344]{BT08} make this argument and cite a test by
\citet{Fren99} which indeed shows that many different codes yield
similar key properties after following the collapse of a dark matter
halo.  In fact, results generally converge in tests that vary the
numerical grid, softening, and/or number of particles
\citep[\eg][]{Powe03,DMS4}, which they would not do if there were a
large element of stochasticity.  \citet{Sell08} also demonstrated
exquisitely reproducible evolution of halo models that were perturbed
by externally imposed bars, in sharp contrast to the results presented
here.

Simulations with active discs of particles, on the other hand, are not
so well behaved.  \citet{SD06} reported some minor differences, and
one major, in a set of experiments using different numerical
parameters but the same file of initial coordinates.  We show here
that simulations with discs can, at least for certain models, exhibit
bi-modally divergent macroscopic results, even between cases that
differ only at the round-off error level.  The reason for this
qualitative difference for discs is because collective instabilities
and vigorous responses develop from particle noise.  Here we identify
a number of distinct causes of stochastic behaviour in discs, and
demonstrate explicitly how the evolution is affected.

We show that the principal sources of divergent behaviour are: (a)
multiple in-plane global modes, (b) swing amplified noise, (c) bending
instabilities, (d) suppression of dynamical friction, and (e) the
truly chaotic nature of $N$-body systems.  We also show that the
distribution of evolutionary paths taken in simulations of
different realizations of the same model varies systematically with
the care taken to set up the initial coordinates of halo particles.

We deliberately choose to illustrate just how large the differences
can be for one particular unstable equilibrium model.  Stochasticity
is present in all simulations and its effects are always noticeable in
those containing discs, but generally variations in the evolution show
less scatter than in the case studied here.  We show that the range of
behaviour is similar in two quite distinct $N$-body codes and
illustrate the sensitivity to differences at the round-off error
level.  We also show that increasing the number of particles does not
reduce the spread of measured properties.

Real galaxies are assembled and evolve in a complicated manner, and
certainly do not pass through a well-constructed axisymmetric,
equilibrium phase that is unstable, although such a model is commonly
used as a starting point of simulations.  The objectives of
experiments of this type are therefore (1) to determine whether
plausible axisymmetric galaxy models are globally stable and (2) to
develop an understanding of the dynamical evolution of models that
form bars and other non-axisymmetric structures.  While we adopt
a model of this type in this paper, its remarkable behaviour has
implications for all simulations of disc-halo models, regardless
of how they were created.

The main part of the paper demonstrates the role of the five
above-named sources of stochasticity in the evolution of disc models.
We also explicitly show the effects of different particle selection
techniques on the robustness of the behaviour.  Stochastic divergence
has been reported elsewhere, but not recognized as an intrinsic aspect
of these models; \eg, \citet{KVCQ} attributed divergent evolution to
inadequate numerical care, whereas stochasticity could be the cause.
Appendix B reports extensive tests that confirm that the results we
report here do not depend on any numerical parameters.

\section{Selection of particles}
\label{selection}
The selection of initial particle positions and velocities of an
equilibrium model requires careful attention.  Random selection of
even many millions of particles will lead to shot noise variations in
both the density and velocity distributions of a model.  Here we
summarize the available techniques to select initial coordinates of
particles, with a focus on disc-halo models.  These methods generally
yield a set of particles that are not specific to any particular
$N$-body code.

\subsection{Selecting from a \DF}
\label{determ}
Jeans theorem requires that an equilibrium model should have a
distribution function (\DF) that is a function of the isolating
integrals (BT08, p.~283).  Thus the best way to realize an equilibrium
set of particles for an initial model is to select from a \DF, when
one is available.

While random selection of particles may be common practice, it
immediately discards a large part of this potential advantage.  One
widely used technique \citep[\eg][]{HBWK,WK07,ZM08,DBS9} is to accept
or reject candidate particles based on a comparison of a random
variable with the value of the \DF\ at the phase-space position of
each particle, which introduces shot noise in the density of particles
in integral space.  The evolution of the simulation will be that of
the selected \DF, not the intended one, and different random
realizations lead to significant variations in the measured
frequencies of the instabilities in the linear regime \citep{Sell83}
and substantial differences in the non-linear regime.  It is therefore
best to adopt a deterministic procedure for particle selection from a
\DF.

A scheme to select particles smoothly in this way, first used in
\citet{Sell83} and described more fully in \citet{SA86}, is summarized
in the Appendix of \citet{DS00}.  We divide integral, generally
$(E,L)$, space into $n$ areas in such a way that $\int\int F dEdL$
over each small area is exactly $1/n$th of the integral over the total
accessible ranges of $E$ \& $L$.  Here $F(E,L)$ is the differential
distribution after integration over the other phase space variables
(BT08, pp.~292, 299).  Requiring that one particle lies within each
area ensures that the selected set of particles is as close as
possible to representing the desired particle density in integral
space.  We choose the precise position of a selected particle within
each area quasi-randomly in order to ensure that the particles do not
lie on an exact raster in integral space.  We describe this scheme as
deterministic selection from the \DF, a term that ignores this minor
random element.

This scheme is readily adapted to select particles of unequal masses
if desired.  To select particles having masses proportional to a
weight function $w(E,L)$, one simply weights the \DF\ by $w^{-1}$,
which automatically adjusts the subdivision of $(E,L)$-space into
areas of equal weighted \DF, as described in \citet{Sell08}.

The phases of the particles around the orbit defined by these
integrals can be selected at random.  We have no evidence that the
choice of radial phase, either for flat discs or for spheres, causes
significant variations in the outcome and we discuss the choice of
azimuthal phases in Section~\ref{qstart} below.

\citet{DS00} describe the similar procedure for 2-integral spheroidal
models.

\subsection{When No Simple \DF\ Is Available}
\label{jeansapp}
Comparatively few useful mass models have known \DF s, and the
realization of an equilibrium set of particles for a general model
presents a significant challenge.  Some authors \citep[\eg][]{SN93}
have simply created a rough $N$-body system, which they then evolve in
the presence of a frozen disc, thereby allowing the halo to relax
towards some nearby equilibrium.

\citet{Hern93} advocates solving the Jeans equations for each
component in the combined potential of all mass components.  His
method is widely used \citep[\eg][]{VK03,Atha03,EZ04,KVCQ}, but the
resulting equilibrium is approximate.

In general, it is better to derive an approximate \DF\ for a spherical
or spheroidal system.  An isotropic \DF\ for a spherical system can
usually be obtained by Eddington inversion (BT08, p.~289), although it
is important to verify that the function is positive for all energies
(which it generally is, for reasonable mass models).

Creating an equilibrium \DF\ for a multi-component system presents a
greater challenge, for which three effective approaches have been
developed.  \citet{RSJK}, \citet{KD95} and \citet{DS00} employ the
method of \citet{PT70} to derive the mass distribution for a halo
having some assumed \DF\ that will be in equilibrium in the presence
of one or more other mass components.  Alternatively, one can use
Eddington's inversion formula for the halo only in the potential of
the combined disc and halo \citep{HBWK}.  A third possibility, as
here, is to start from a known spherical halo with a known \DF\ and
compress it by adding a disc and/or a bulge using Young's (1980)
method \citep[see][]{SM05}, and then to select particles from the
compressed \DF.  Even though the last two methods use only the
monopole term for the disc, all three methods yield a spheroidal
system that is close to detailed equilibrium everywhere.

In general, it is more difficult to construct a good equilibrium for a
disc component.  The circular speed in the disc mid-plane as a
function of radius is determined by the total mass distribution and,
commonly, one specifies $Q(R)$ \citep{Toom64} to determine the radial
velocity spread at each radius.  The Jeans equations in the epicycle
approximation (BT08, p.~326) generally yield a poor equilibrium except
when the radial dispersion is a small fraction of the circular speed,
and the asymmetric drift formula may have no solution near the centres
of hot discs.  \citet{Shu69} describes an approximate \DF\ for a warm
disc with a given radial velocity dispersion that we, and
\citet{KD95}, have found to be quite serviceable.  Again in cases
where the radial velocity dispersion stretches the validity of the
epicycle approximation, radial gradients can lead to a disc surface
density after integration over all velocities that differs slightly
from that specified, as shown in Section~\ref{model}.

The vertical structure of an isothermal stellar sheet is given by the
formulae developed by \citet{Spit42} and \citet{Camm50}, and BT08
(p.~321) describe a generalization of the in-plane \DF\ to include
this feature, which they describe as the Schwarzschild \DF.  The
Spitzer-Camm formulae assume full Newtonian gravity and no radial
density or dispersion gradient.  Force softening has an increasingly
detrimental effect on the vertical balance as the ratio of disc
thickness to softening length is reduced, we therefore prefer to
construct a vertical equilibrium from the 1D vertical Jeans equation
in the actual force field of the softened disc potential, which leads
to a better equilibrium.

\subsection{Quiet Starts}
\label{qstart}
The quiet start technique is a valuable addition to the set up process
only when the model has a few vigorous, large-scale instabilities,
such as arise in a cool, massive disc with a rotation curve that rises
approximately linearly from the centre.  It is of little help when
linear stability theory predicts the model to be responsive but
(almost) stable \citep[\eg][]{Sell89,SE01}.  In these latter cases,
collective responses to residual noise grow more vigorously than any
global modes, and the particle arrangement randomizes quickly.

For a quiet start, one reproduces each selected master particle
multiple times in a symmetrical arrangement, with image particles
having the identical radius and velocity components in polar
coordinates.  We restrict the meaning of the phrase ``quiet start'' to
this symmetrical arrangement of particles -- \ie\ a quiet start can be
used no matter how the coordinates of the master particles are
selected.  Conversely, a ``noisy start'' means only that azimuthal
coordinates are selected at random, again independent of how the
master particles are selected.  The procedures for discs and
spheroidal components differ slightly.

For discs, we place image particles at the corners of an almost
regular polygon in 2D, centred on the model centre.  The polygon is
not exactly regular because we nudge the particles away from exact
$n$-fold symmetry by a random fraction of a small angle, typically
$0.02^\circ$.  When the disc has a finite thickness, the polygon must
be duplicated with a second on the opposite side of the mid-plane for
which both the vertical position $z$ and velocity $v_z$ of every
particle in each of the two polygons have opposite signs.

When the force-determination method is based around an expansion in
sectoral harmonics that is truncated at low order, $m_{\rm max}$, and
the number of sides to the polygon $n \geq 2m_{\rm max}+1$, azimuthal
forces in the initial model are much lower than would arise from
particle shot noise -- hence the label ``quiet start''.

\begin{figure}
\centerline{\includegraphics[width=.8\hsize]{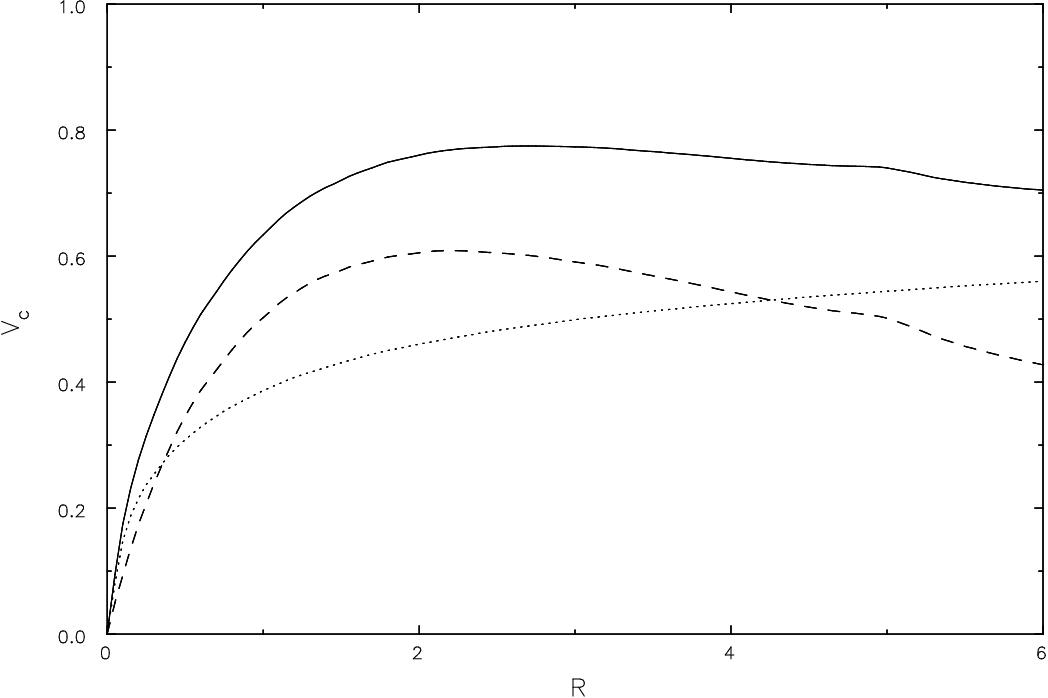}}
\caption{The inner rotation curve of our standard model
(solid).  The separate contributions of the disc (dashed) and halo
(dotted) are also shown.}
\label{rchern}
\end{figure}

We have not tried quiet starts for other force methods, but they could
still offer a significant advantage provided that the number of
corners adopted for the polygon exceeds the azimuthal order of all the
strong instabilities and non-axisymmetric responses (Section~\ref{swamp}) by
at least a factor two.

\begin{figure*}
\centerline{\includegraphics[width=.57\hsize,angle=270]{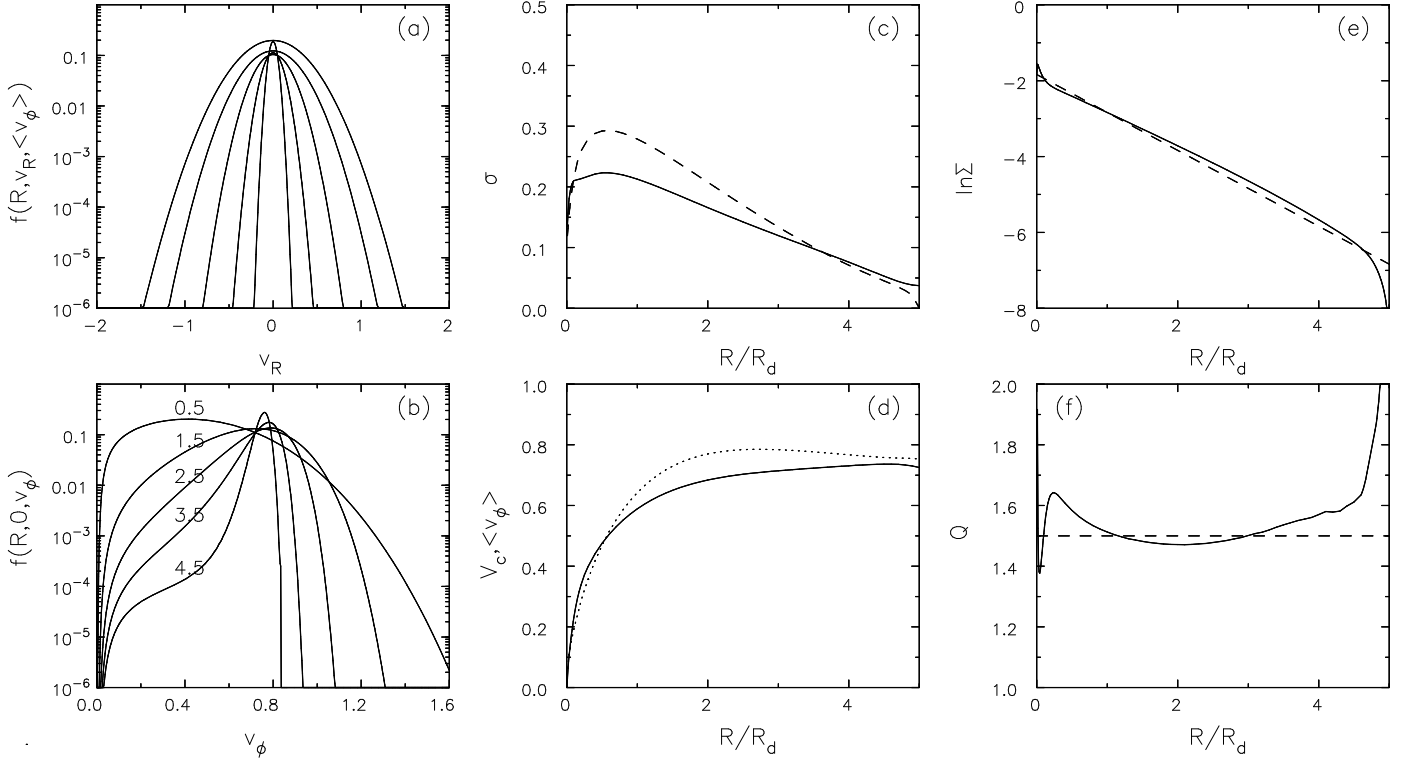}}
\caption{Details of the approximate \DF\ for the disc.  Panels (a) and
(b) show respectively the variation of $f$ with radial velocity and
azimuthal velocity at five different radii.  Panel (c) shows the
radial variations of the rms azimuthal speed ($v_\phi$ solid) and
radial speed ($v_r$ dashed), (d) compares the circular speed (dotted)
with the mean $v_\phi$ (solid) to illustrate the asymmetric
drift. Panels (e) and (f) compare respectively the actual surface
density and $Q$ profiles (solid) with the desired profiles (dashed).
The \DF\ does not reproduce these curves perfectly, but the departures
are minor.}
\label{plotdf}
\end{figure*}

We adopt a similar procedure for spheroidal components, except that we
create image particles by rotating the initial position and velocity
vectors using the usual rotation matrix for the adopted set of Euler
angles \citep[\eg][p.~199]{Arfk85}.  The set of Euler angles used
creates an $n$-fold rotationally symmetric set of particles, which is
also reflection symmetric about the mid-plane, and has zero net
momentum with a centre of mass at the model centre; each master
particle is therefore inserted $2n$ times.  It is reasonable to adopt
$n \ga 4$.

\section{Models}
Here we describe all the various galaxy models we use in this paper.

\subsection{Standard Galaxy Model}
\label{model}
Our standard model is a composite disc-halo system with the rotation
curve shown in Fig.~\ref{rchern}.  The two mass components are an
exponential disc and a compressed, strongly truncated, Hernquist halo.

The initial surface density of the disc has the usual exponential form
\begin{equation}
\Sigma(R) = {M_d \over 2\pi R_d^2}e^{-R/R_d},
\end{equation}
where $M_d$ is the nominal disc mass.  We truncate the disc at
$R=5R_d$, leaving an active disc mass of $\approx 0.96 M_d$.  The disc
particles are set in orbital motion with a radial velocity spread so
as to make Toomre's $Q=1.5$.  For most models, we determine the
approximate equilibrium velocities by solving the Jeans equations in
the epicycle approximation as described in Section~\ref{jeansapp}.

In some cases we adopt Shu's approximate \DF\ instead, and select disc
particles deterministically from it.  Properties of the \DF\ and the
radial variations of the low-order velocity moments are shown in
Fig.~\ref{plotdf}.  While the radial velocity distributions are
nicely Gaussian, the azimuthal velocity distributions (\ref{plotdf}b)
are markedly skewed.  This aspect, and the departures of the surface
density and $Q$ profiles from the desired values all decrease for
models with less dominant discs or with lower values of $Q$.

For fully 3D simulations, the density profile normal to the disc
plane is Gaussian, with a constant scale height of $0.05R_d$ and
appropriate vertical velocities in the numerically determined vertical
force profile.

We construct a halo in equilibrium with the disc in the following manner.
We start from the initial density profile suggested by \citet{Hern90}
\begin{equation}
\rho_0(r) = {M_h r_s \over 2\pi r(r_s + r)^3},
\label{Hernquist}
\end{equation}
which has total mass $M_h$ and scale radius $r_s$, with the isotropic
distribution function (\DF) also given by Hernquist.  We strongly
truncate this halo by eliminating all particles with enough energy to
reach $r > 2r_s$, causing the density to taper gently to zero at this
radius, and an actual halo mass of $\approx 0.25M_h$.  Since most of
the discarded mass is at large radii, there is little change to the
central attraction at $r<2r_s$ and the model remains close to
equilibrium.

For our standard model, we choose $r_s = 40R_d$ and set $M_h = 80 M_d$
so that the halo mass is approximately 19 times that of the disc.  We
then employ the halo compression algorithm described by \citet{SM05}
to compute a new, mildly anisotropic, \DF\ for the compressed halo
that results from including the above disc.  The rotation curve,
Fig.~\ref{rchern}, shows that the disc dominates the central
attraction over most of the inner part and the total rotation curve is
approximately flat at large radii.

We adopt a system of units such that $G = M_d = a_d = 1$, where $G$ is
Newton's constant, $M_d$ is the mass of the untruncated disc, and
$a_d$ is the length scale for the type of disc adopted.  Therefore
distances are in units of $a_d$, masses are in units of $M_d$, one
dynamical time $\tau = (a_d^3/GM_d)^{1/2}$, and velocities are in
units of $\hat v = (G M_d / a_d)^{1/2} \equiv a_d / \tau$.  One
possible scaling to physical units is to choose the dynamical time to
be $10\;$Myr and $a_d = 3\;$kpc, which implies $M_d = 5.98 \times
10^{10}\;{\rm M}_\odot$.  The velocity unit $\hat v =
293\;$km~s$^{-1}$, and the peak circular speed in Fig.~\ref{rchern}
is approximately $235\;$km~s$^{-1}$.

We also present results for two other disc-halo models for which we
choose $r_s = 30R_d$ and $r_s = 50R_d$, \ie\ that bracket our standard
case.  The more extended halo leads to a more dominant disc, while the
disc is less dominant in the more concentrated halo.

\begin{figure}
\centerline{\includegraphics[width=.8\hsize]{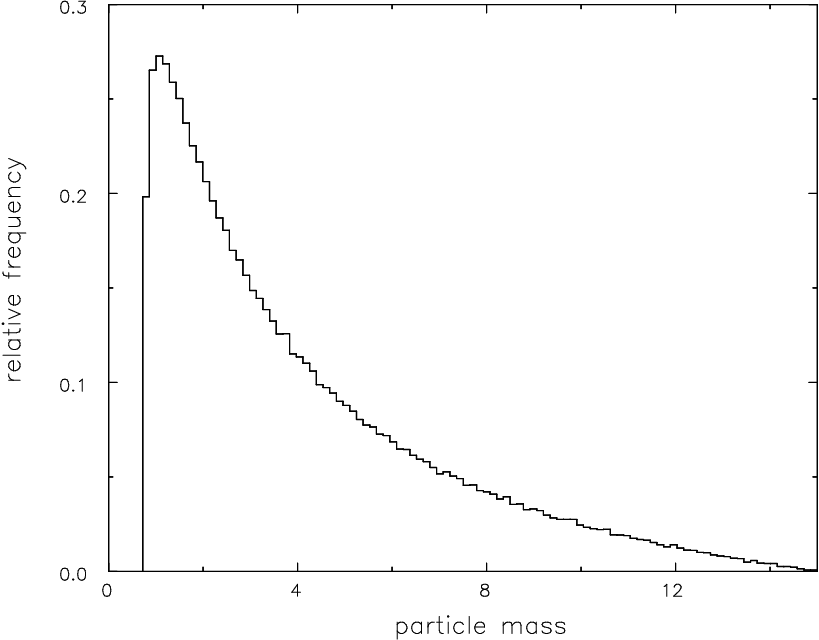}}
\caption{The frequency distribution of halo particle masses, in units
of the disc particle mass.}
\label{hist}
\end{figure}

We select halo particles from the compressed \DF\ using the smooth
procedure summarized in Section~\ref{determ}, with the weight function for
particle masses being $w(L) = 0.5 + 20L$, where $L = |\bL|$ is the
total specific angular momentum.  All disc particles have equal
masses, but the masses of halo particles range from 0.7 to 14.6 times
the mass of the disc particles.  Fig.~\ref{hist} shows the frequency
distribution of halo particle masses.

As a result of this careful procedure, both the disc and halo
components are very close to equilibrium in the combined potential and
the initial ratio of kinetic to the virial of Clausius (measured from the
particles) is $T/|W|=0.498$.  At the same time, the phases of the
particles in their carefully selected orbits are chosen at random, so
that the model indeed starts from the usual level of shot noise
resulting from the random locations of the particles.

\subsection{Isochrone Disc}
We also present results using the isochrone disc with no halo.  The
potential (BT08, p.~65) has a simple form
\begin{equation}
\Phi(R) = -{GM_d \over a} \left[x + (1 + x^2)^{1/2}\right]^{-1},
\end{equation}
while the surface density is
\begin{equation}
\Sigma(R) = {M_d a \over 2\pi r^3} \left\{ \log\left[ x + (1 +
x^2)^{1/2} \right] - {x \over (1 + x^2)^{1/2}} \right\}.
\end{equation}
Here $a$ is a length scale, and $x=r/a$; note $\Sigma(0)=M_d/(6\pi
a^2)$.  \citet{Kaln76} describes a convenient family of \DF s
characterized by a parameter $m_K$; we refer to each model as the
isochrone/$m_K$ disc.  He \citep{Kaln78} also presents some
preliminary results for the normal modes, which were confirmed in
simulations \citep{ES95}.  The local stability parameter
\citep{Toom64} for the isochrone/5 disc has an near constant value of
$Q \simeq 1.6$, and is $Q \simeq 1.2$ for the isochrone/8 model.

\begin{table}
\caption{Numerical parameters for our standard runs}
\label{params}
\begin{tabular}{@{}lrr}
                   & Cylindrical grid       & Spherical grid \\
\hline
Grid size          & $(N_R,N_\phi,N_z)\quad$ \\
                   &  $ = (127,192,125)$    & $n_r = 500$ \\
Angular compnts    & $0\leq m \leq 8$       & $0 \leq l \leq 4$ \\
Outer radius       & $6.076R_d$             & $80R_d$ \\
$z$-spacing        & $0.01R_d$ \\
Softening rule     & cubic spline           & none \\
Softening length   & $\epsilon = 0.05R_d$  \\
Number of particles & 500\,000              & 2\,500\,000 \\
Equal masses       & yes                    & no (see Fig.~\ref{hist}) \\
Shortest time step & $0.0125(R_d^3/GM)^{1/2}$ & $0.0125(R_d^3/GM)^{1/2}$ \\
Time step zones    & 5 & 5 \\
\hline
\end{tabular}
\end{table}

\section{Results}
\label{illust}
\begin{figure}
\includegraphics[width=.57\hsize,angle=270]{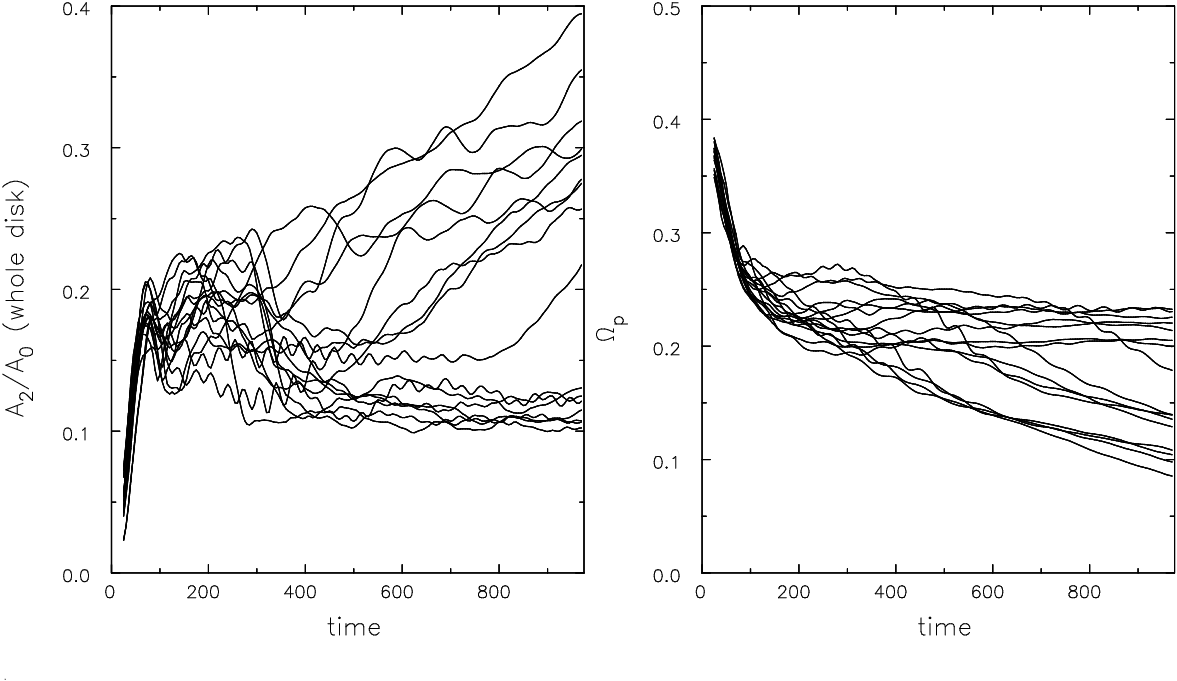}
\caption{Evolution of the amplitude (left) and pattern speed (right)
of the bar in 16 runs with different random seeds for the disc
particle coordinates, run using Sellwood's (2003) hybrid code.  The
tiny differences in the initial models lead to a remarkably wide range
of properties of the bar at late times.}
\label{basichyb}
\smallskip
\includegraphics[width=.57\hsize,angle=270]{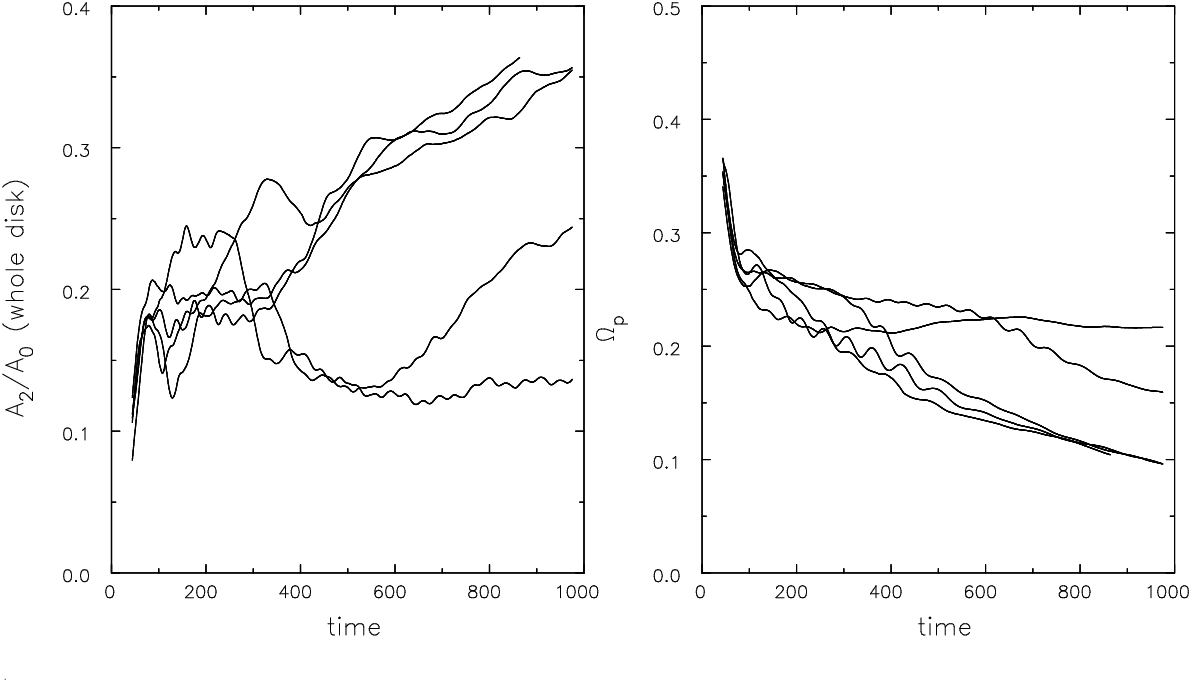}
\caption{Evolution of 5 runs with different random seeds for the disc
particle coordinates, run using \PK\ with $\epsilon = 0.05R_d$.}
\label{pkdgrav}
\end{figure}

We begin by showing just how much variation can occur.  We first
present the evolution of our standard disc/halo model whose rotation
curve is shown in Fig.~\ref{rchern}.  Note that the disc equilibrium
in these models is set by solving the Jeans equations, while the halo
particles are selected deterministically from a \DF.
Fig.~\ref{basichyb} shows results from 16 separate runs with Sellwood's
(2003) hybrid grid code using fixed numerical parameters, given in
Table~\ref{params}, but with different random seeds for the initial
coordinates of the disc particles only.  We plot the evolution of both
the amplitude and pattern speed of the bar, measured as described in
Appendix A.  Even though the initial particles are selected from the
same distributions, with different random seeds for the disc only, the
amplitude evolution differs greatly from run to run and there is
considerable spread in the evolution of the pattern speed.

In order to demonstrate immediately that the scatter in
Fig.~\ref{basichyb} is not a numerical artefact of our grid code,
Fig.~\ref{pkdgrav} shows the results of a similar test with 5 runs
using the tree code \PK\ \citep{Stad01} using an opening angle $\theta
= 0.7$.  \PK\ is a multi-stepping code, with time steps refined such
that $\delta t = \Delta t/2^n < \eta (\epsilon/a)^{1/2}$, where
$\epsilon$ is the softening and $a$ is the acceleration at a
particle's current position.  We use base time step $\Delta t = 0.01$
and $\eta = 0.2$, which gives identical time steps for all particles.
The results show a comparable spread in the evolution of both the
amplitude and pattern speeds.  Results from the two codes with
identical initial coordinates for all the particles do not compare in
detail.  For this problem, the tree code runs about 37 times more
slowly than Sellwood's (2003) grid code; we therefore use it only for
this cross check.

The gross qualitative behaviour of all the models in
Figs.~\ref{basichyb} \& \ref{pkdgrav} is similar at first.  The bar forms at
similar times with similar pattern speeds, though the initial peak
amplitude varies by about $\sim 25\%$.  The evolution thereafter
further diverges, notably with increasingly large differences in the
bar amplitude.  Steep declines in the bar amplitude in the interval
$200 \la t \la 400$ are generally associated with buckling events
\citep[\eg][]{RSJK}, but the timing of these events varies
considerably.  At late times in Fig.~\ref{basichyb}, the bar
amplitude rises steadily in 9/16 simulations, although starting from
different times in each case, while it stays low (over the time
interval shown) in the remaining 7.

It is more encouraging to note that the rate of decrease of the bar
pattern speed does correlate with the bar amplitude; strong bars are
more strongly braked by halo friction, as expected.  Furthermore,
continued amplitude growth of bars that are strongly braked has been
reported previously \citep[\eg][]{Atha02}.

\begin{figure}
\centerline{\includegraphics[width=.8\hsize]{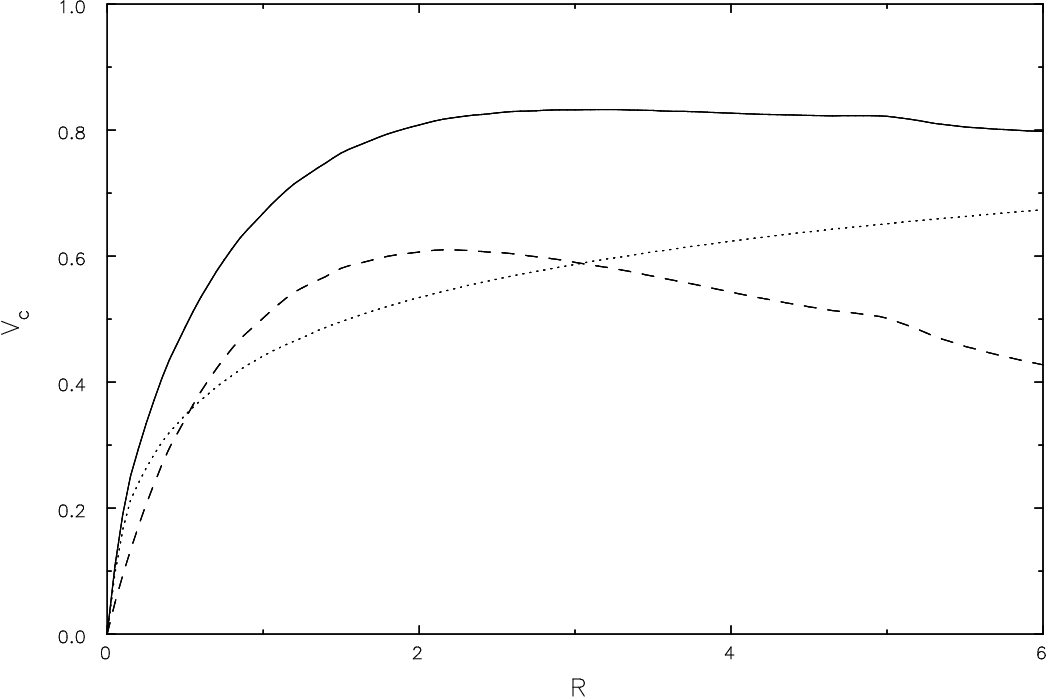}}
\medskip
\centerline{\includegraphics[width=.8\hsize]{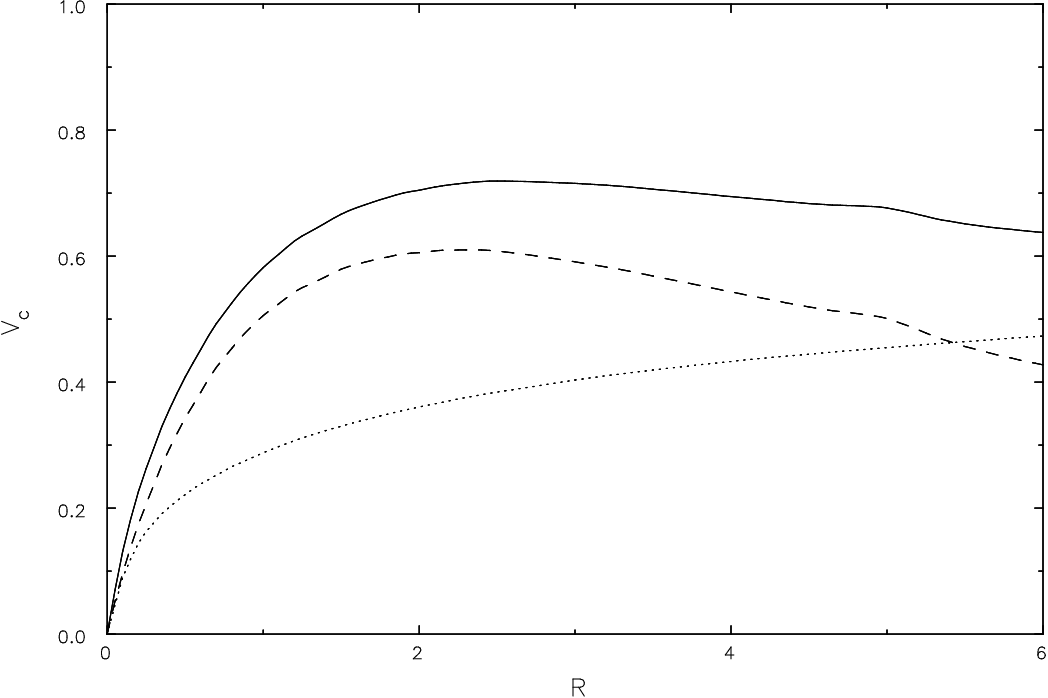}}
\caption{The inner rotation curves of models with (above) a slightly
more domination halo and (below) a slightly more extended halo.  The
line styles mean the same as in Fig.~\ref{rchern}.  The behaviour
of these models is shown in Figs.~\ref{newhalo} \& \ref{exthalo}}.
\label{rcother}
\end{figure}

\begin{figure}
\includegraphics[width=.57\hsize,angle=270]{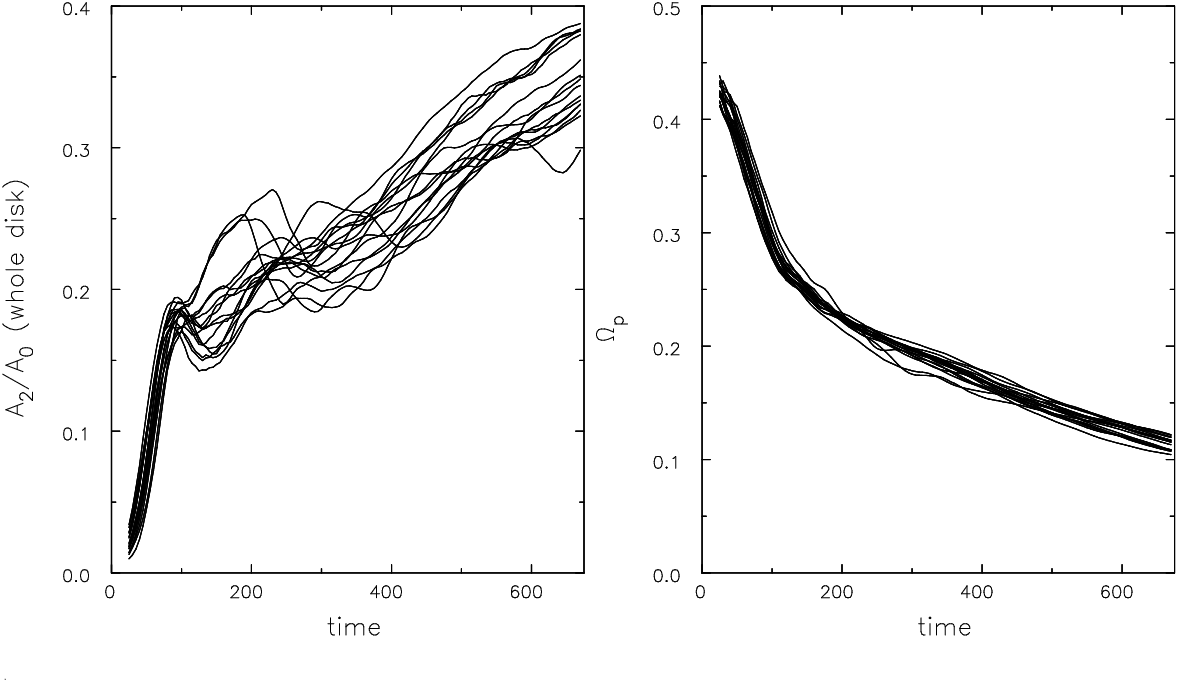}
\caption{Evolution of a set of models with a more dominant halo than
those shown in Fig.~\ref{basichyb}.  The initial rotation curve is
shown in the upper panel of Fig.~\ref{rcother}.}
\label{newhalo}
\vspace{0.3cm}
\includegraphics[width=.57\hsize,angle=270]{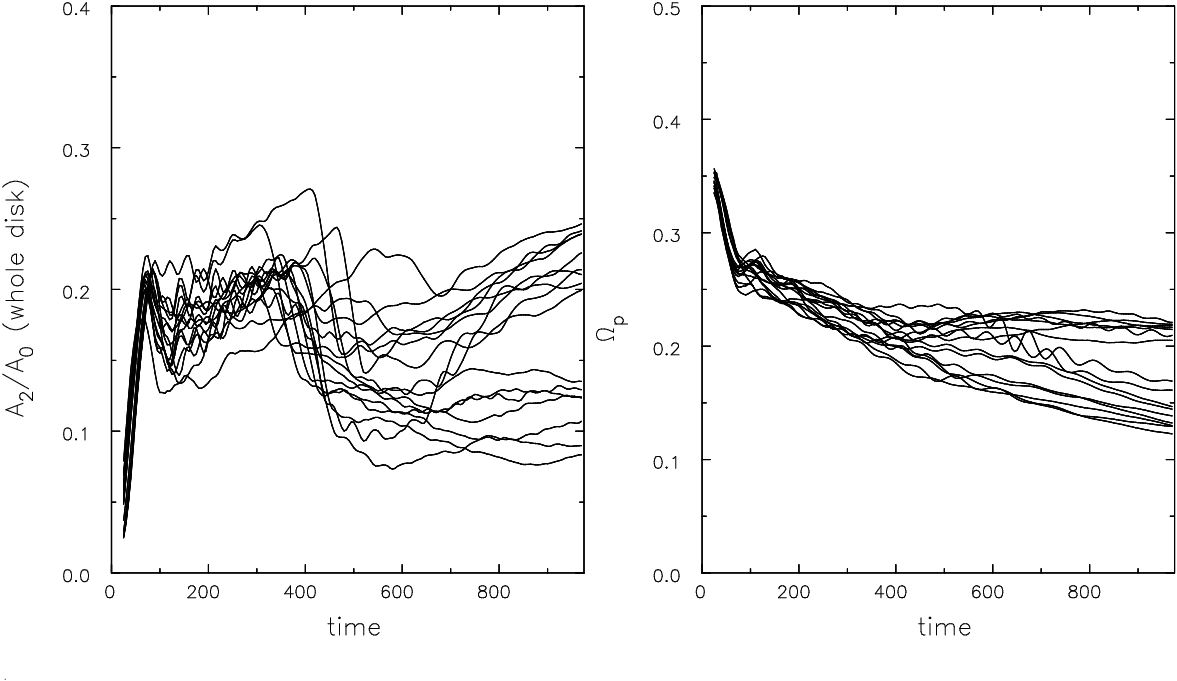}
\caption{Evolution of a set of models with a less dominant halo than
those shown in Fig.~\ref{basichyb}.  The initial rotation curve is
shown in the lower panel of Fig.~\ref{rcother}.}
\label{exthalo}
\vspace{0.3cm}
\includegraphics[width=.57\hsize,angle=270]{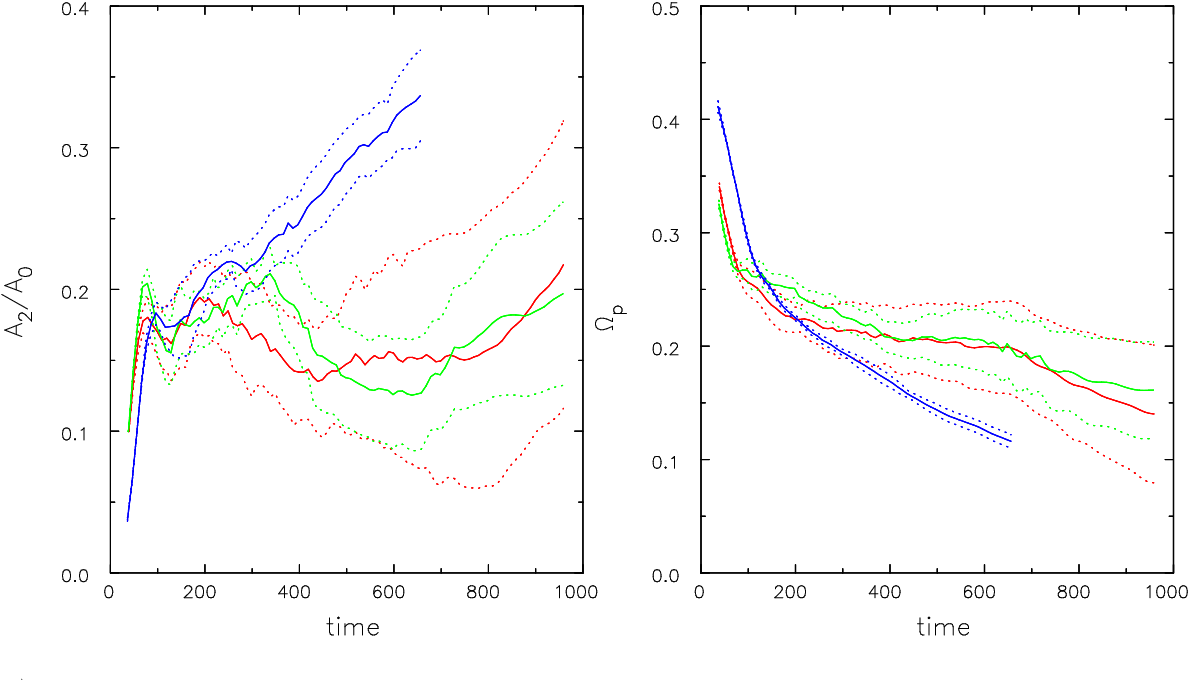}
\caption{Comparison of the estimated means (solid lines) and
$\pm1\sigma$ scatter (dotted curves) in the three different haloes
shown in Figs.~\ref{basichyb} (red), \ref{newhalo} (blue), \&
\ref{exthalo} (green).}
\label{threehalos}
\end{figure}

\subsection{Divergence at Late Times}
\citet{DBS9} report a similar study of bar-unstable disc-halo models,
which also reveal large amplitude differences in the short term.
However, they stress that the long-term evolution of their simulations
is reproducible, in contrast to our finding.

Fig.~\ref{newhalo} shows that we confirm their conclusion for a
different model with a slightly more dominant halo; the evolution of
both the bar amplitude and pattern speed shows much less scatter than
is seen in Fig.~\ref{basichyb}.  All cases show a steady rise in bar
amplitude after the buckling event, although the curves for the
different realizations during this stage of the evolution are offset
in time, as also found by Dubinski \etal

Fig.~\ref{exthalo} shows results from a third model with a more
dominant disc.  The amplitude evolution in this model is again
bi-modal, rising steadily at late times in half the cases, although
not by as much as in our standard case (Fig.~\ref{basichyb}).  The
rotation curves of both these models are shown in
Fig.~\ref{rcother}.

The late rise in bar amplitude occurs, if at all, only in models with
live haloes and is associated with frictional braking.  It is natural
that frictional braking should be stronger when the halo is more
dominant.  In our standard model (Fig.~\ref{basichyb}), and in the
more dominant disc case (Fig.~\ref{exthalo}), the large late-time
differences arise because strong friction kicks in in some cases but
not in all.  We argue in Section~\ref{dynfr} that the reason for these
differences is the existence of adverse gradients in the halo \DF,
which can inhibit friction \citep{SD06}.  Whatever the cause, it is
clear from these two sets of runs that onset of friction and steady
bar growth at late times depends on comparatively minor differences in
the earlier evolution caused by the different random seeds.

In order to quantify the scatter, we compute the bi-weight estimate
\citep{Beers} of the mean and dispersion of the measurements
throughout all sets of experiments.\footnote{Their algorithm assumes
the data to be unimodal with a few outliers, which is manifestly not
the case in our data at late times.}  Since bar growth is shifted
slightly in time in the different runs shown in
Figs.~\ref{basichyb}, \ref{newhalo}, and \ref{exthalo}, we apply a
small time offset to the evolution of both quantities in order to
ensure that the evolution coincides as the relative bar amplitude
grows through 0.1, before computing the mean and scatter from each
set.  Fig.~\ref{threehalos} shows the time evolution of the means
and scatter of the bar amplitude and pattern speed for all three
haloes.  It is clear that the stochastic spread is greatest for our
standard halo (red lines), less for the less dominant halo (green
lines) and least for the more dominant halo (blue lines).

\begin{figure}
\includegraphics[width=.57\hsize,angle=270]{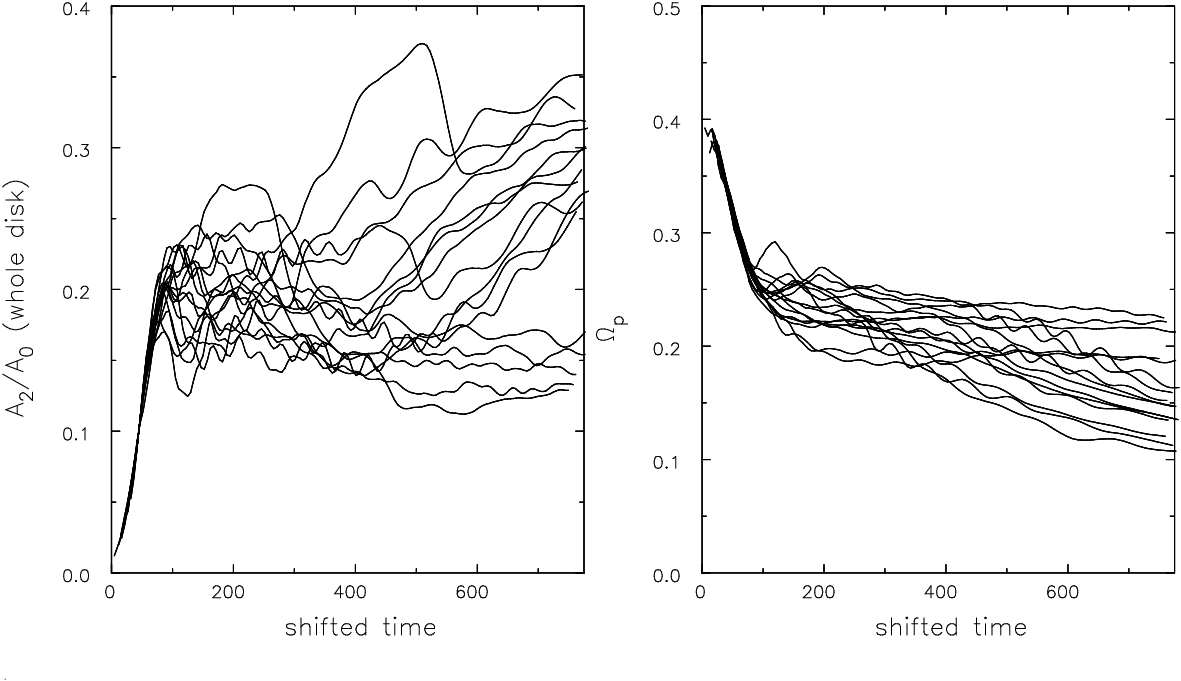}
\caption{Evolution of a set of 16 runs of our standard model that used
a more careful disc set up procedure.}
\label{expdfdetn}
\medskip
\includegraphics[width=.57\hsize,angle=270]{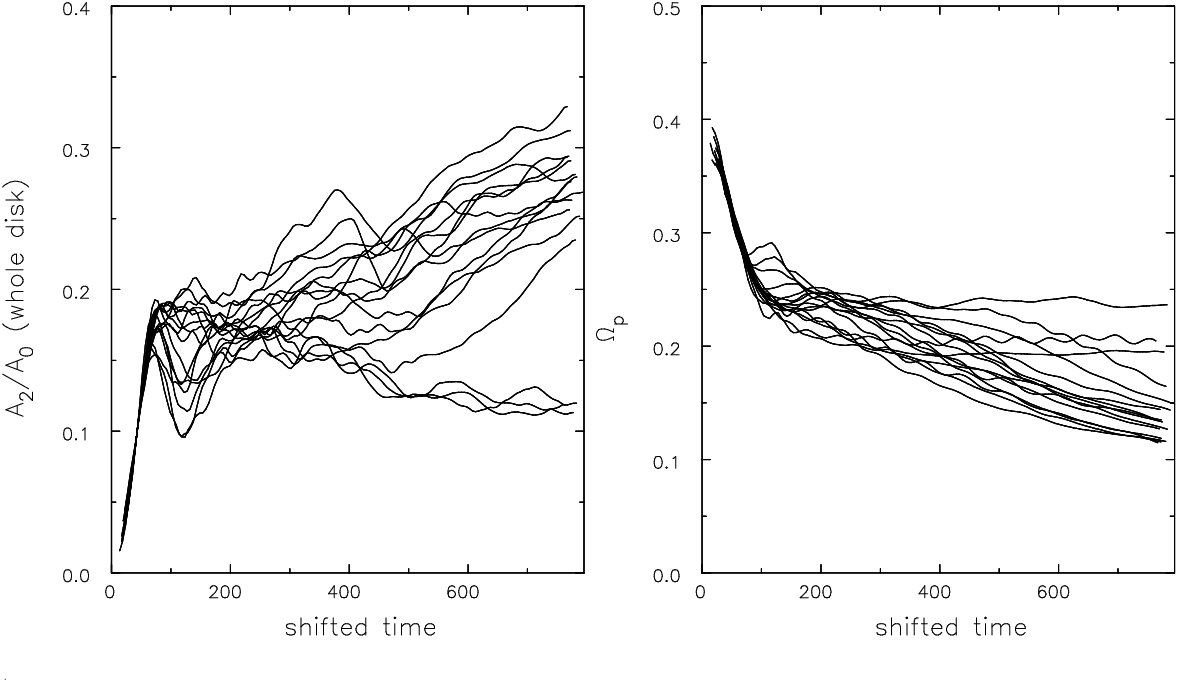}
\caption{Evolution of a set of 16 runs that used Hernquist's Jeans
equation procedure to set up an approximate equilibrium for the halo
particles.  The bar amplitude grows at late times and the pattern
decreases in all but three of these cases.}
\label{ranhal}
\end{figure}

\subsection{Particle selection}
\label{pselect}
Fig.~\ref{expdfdetn} shows the consequence of selecting disc
particles in a deterministic manner from an approximate \DF\ as
described in Sections~\ref{jeansapp} \& \ref{model}.  This procedure still
has a random element when choosing the precise values of $E$ \& $L_z$
within each sub-area, and the simulations have noisy starts because we
randomly select the radial and azimuthal phases of the particles.  The
16 different runs used different random seeds and are to be compared
with those shown in Fig~\ref{basichyb}, for which disc particle
velocities were selected from Gaussians whose widths were estimated
from the Jeans equations in the epicycle approximation.  There is no
significant improvement, and in this case 6/16 runs have not slowed
much by $t=800$.

The consequences of selecting {\it halo\/} particle velocities from
Gaussians whose widths are determined from the Jeans equations
\citep{Hern93}, are shown in Fig.~\ref{ranhal}.  With this more
approximate halo equilibrium we see that all but 3/16 bars grow and
slow.  The non-slowing fraction was 5/16 in a similar set of
experiments (not shown) in which the halo particles were selected from
the \DF\ by the accept/reject method, instead of deterministically for
Fig.~\ref{basichyb}.

Thus we find a weak trend in these results with the quality of the
different halo set-up procedures.  The fraction of bars that do not
experience strong friction rises to almost half when we use the most
careful set-up procedure we have been able to devise for the halo,
whereas use of the density profile to choose radii and Jeans equations
to set halo velocities results in a large majority (13/16) of bars
that experience strong friction (Fig.~\ref{ranhal}).  This trend is
also consistent with the weak dependence on halo particle number
reported in Appendix B, where we find that the larger the halo
particle number, the smaller the fraction of bars that slow.  We also
find a larger fraction of slowing bars when we use equal mass
particles.  These results hint that still larger calculations that are
set up with extreme care may evolve in a consistent manner independent
of the random seed, but we have been unable to demonstrate this.

\begin{figure}
\includegraphics[width=.57\hsize,angle=270]{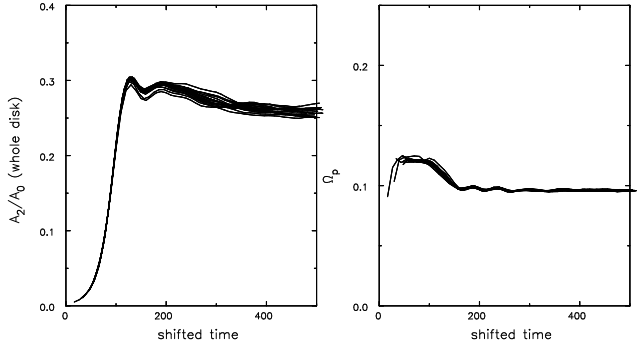}
\caption{Evolution of the bar starting from 16 different selections of
particles from the same DF of the isochrone/5 disc.}
\label{isoc5dfdetn}
\end{figure}

\begin{table}
\caption{Numerical parameters for our 2D simulations}
\label{pars2d}
\begin{tabular}{@{}lrr}
                   & Isochrone & Standard model\\
\hline
Grid $(N_R,N_\phi)$ & (180,256)        & (170,256) \\
Sectoral harmonics  & $0\leq m \leq 8$  & $0\leq m \leq 8$   \\
Outer radius       & $3.995a$ & $6.23R_d$          \\
Softening rule     & Plummer & Plummer \\
Softening length  $\epsilon$ & $0.05a$   & $0.1R_d$  \\
Number of particles & 500\,000 & various \\
Equal masses       & yes   & yes   \\
Shortest time step & $0.025$& $0.0125$  \\
Time step zones    & 1 & 3 \\
\hline
\end{tabular}
\end{table}

\section{Sources of stochasticity}
In this section, we describe and illustrate five sources of
stochasticity, four of which contribute to the large scatter just
described.

\subsection{A Reproducible Result}
We start from a simple unstable disc model for which the outcomes of
simulations do not diverge with different random selections of initial
particles.  Fig.~\ref{isoc5dfdetn} shows results from noisy start
simulations in 2D of an isochrone/5 disc, in which $Q \simeq 1.6$;
numerical parameters are given in Table~\ref{pars2d}.  The different
curves come from separate simulations with different selections of
particles from the {\em same} \DF, using the ``deterministic''
procedure described in Section~\ref{selection}.  The small scatter in the
bar amplitude at late times can be further reduced by restricting
disturbance forces to the $m=2$ sectoral harmonic only.

\begin{figure}
\includegraphics[width=.57\hsize,angle=270]{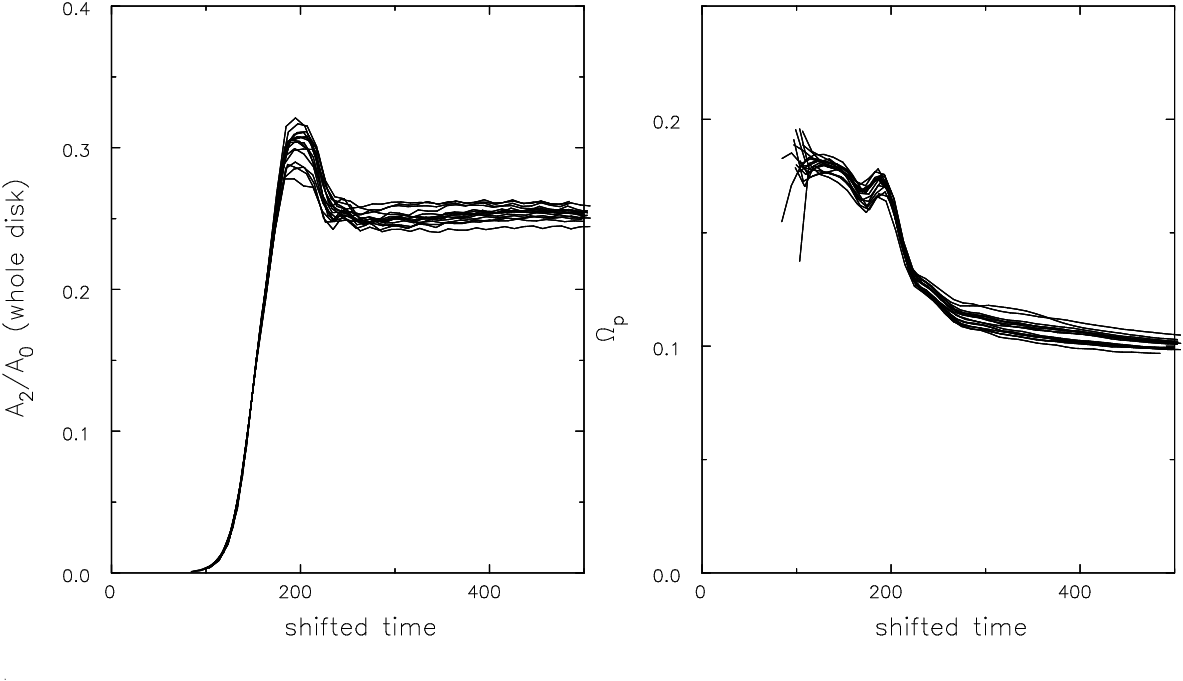}
\caption{The time evolution of the bar amplitude and pattern speed in
a quiet start isochrone/8 disc in which $Q \simeq 1.2$.  Note the
somewhat larger spread compared with that shown in
Fig.~\ref{isoc5dfdetn}.}
\label{isoc8dfdetq}
\vspace{0.25cm}
\includegraphics[width=.57\hsize,angle=270]{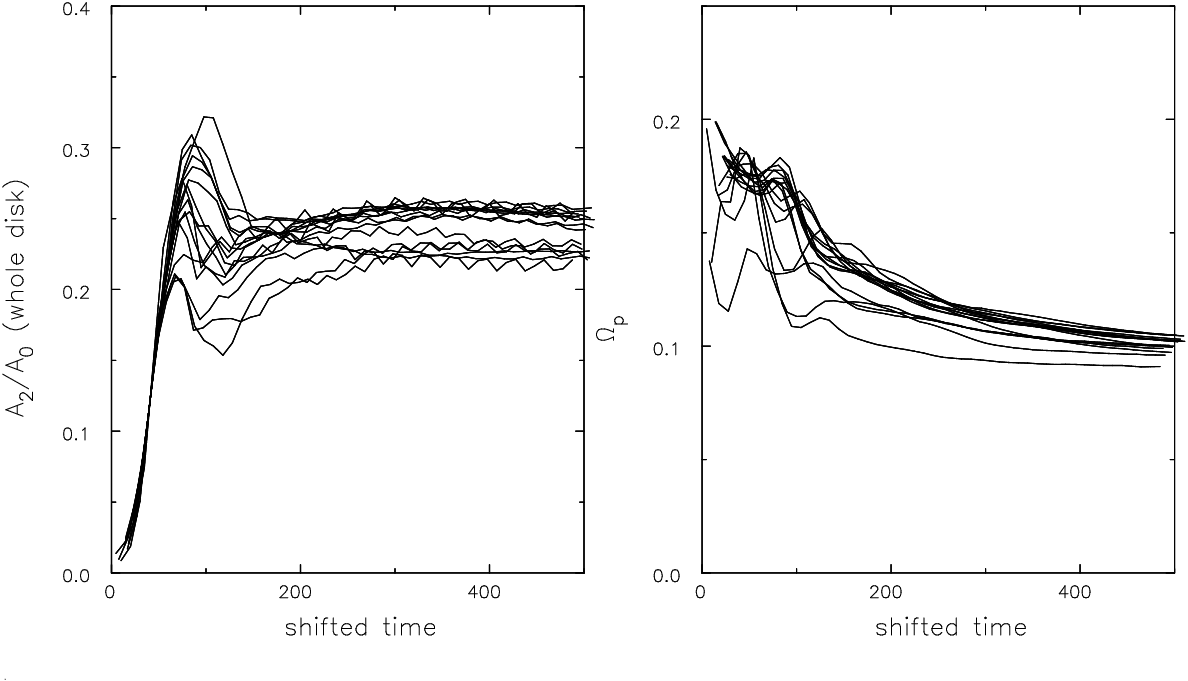}
\caption{Evolution of the bar in a noisy start isochrone/8 disc in which
particles are drawn from the same \DF\ as was used for
Fig.~\ref{isoc8dfdetq}.}
\label{isoc8dfdetn}
\end{figure}

\subsection{Multiple Modes}
\label{multimodes}
Most unstable disc models support a large set of
small-amplitude, unstable modes having a wide range of growth rates
\citep[\eg][]{Toom81,Jala07}.  These linear modes, even those
with the same angular periodicity, grow independently for as long as
all disturbance amplitudes remain small.  If the seed amplitudes of
all modes are low, the first to saturate will be the most rapidly
growing.  In most unstable discs, the fastest growing mode is
generally the simplest, or fundamental, mode that is usually dubbed
the bar mode.  But if the growth rate of the bar mode does not exceed
that of the next most vigorous mode by a large enough margin for some
seed amplitudes, then both may have comparable amplitude when one
saturates.  The consequence of two or more modes reaching large
amplitude at similar times but with random phases can lead to
constructive or destructive interference in the measured amplitudes as
the ``bar'' saturates.  Non-linear effects then cause such differences
to persist.

We use the slightly cooler $m_K=8$ isochrone disc to demonstrate this
behaviour explicitly and, to avoid additional complications, we
restrict disturbance forces to those arising from the $m=2$ sectoral
harmonic only.  Figs.~\ref{isoc8dfdetq} \& \ref{isoc8dfdetn}
illustrate the dependence of the outcome on the initial noise
amplitude.  The quiet start simulations in Fig.~\ref{isoc8dfdetq}
are good enough that the growth rates of the two most rapidly growing
$m=2$ modes can be estimated by fitting to data from the
extensive period of evolution before growth ends \citep[\eg][]{SA86}.
We find the growth rate of the second mode to be some 85\% of that of
the bar mode and that its amplitude (peak $\delta \Sigma/\Sigma$)
can be within a factor of a few of the dominant mode as the bar
saturates.  The consequence is a slight increase in the scatter of the
later bar amplitudes in this case compared with the case for the
hotter disc shown in Fig.~\ref{isoc5dfdetn}.

The mild scatter in Fig.~\ref{isoc8dfdetq} requires a quiet start,
which decreases the seed amplitude of all non-axisymmetric
disturbances that grow for $\sim 100$ time units before the rising
amplitudes even become discernible in the figure.  The much larger
seed amplitudes when noisy starts are used do not allow the dominant
mode to outgrow all others before saturation, with the consequences
illustrated in Fig.~\ref{isoc8dfdetn}.  The same sets of particles
were used as for the results shown in Fig.~\ref{isoc8dfdetq}, but we
placed the image particles at random azimuths, instead of evenly.  The
period of rising amplitude is too short to allow more than very rough
measurments of the growing modes, but it is clear that multiple
unstable modes having comparable growth rates are seeded at large
initial amplitudes by the shot noise.  With such high seed amplitudes,
there is not enough time for the most rapidly growing mode to outgrow
the others, which therefore leads to very substantial variations in
the final bar amplitudes.  Note that this did not happen in the warmer
disc (Fig.~\ref{isoc5dfdetn}), which also used a noisy start, since
in that case all growth rates are lower, while the growth rate of the
dominant bar mode exceeds that of all others by a larger margin.

Notice also that not only is there greater scatter in both the bar
amplitude and pattern speed in Fig.~\ref{isoc8dfdetn}, but both
quantities scatter to lower values.  We find indications that runs
having lower pattern speed have the more dominant second mode.  The
fundamental bar mode, when it has time to outgrow the second mode,
peaks at a greater amplitude and then relaxes back to lower value, as
always happens in Fig.~\ref{isoc8dfdetq}.  But when the second mode
is competitive, the bar amplitude generally has a lower initial peak,
and may even rise subsequently.

\subsection{Swing-amplified noise}
\label{swamp}
Our standard model is more complicated than the isolated isochrone
disc.  In particular, the inner rotation curve (Fig.~\ref{rchern})
rises steeply where the halo density cusp dominates.  Recall that a
mode is a standing wave oscillation of the system, which can be
neutral, growing, or decaying.  The dominant linear global modes,
known as cavity modes, in bar unstable discs are standing waves
between the centre and corotation that must have a high enough pattern
speed to avoid any inner Lindblad resonances
\citep[][pp.~508-518]{Toom81,BT08}.  The consequence of a steeply
rising rotation curve is to make the maximum of the function $\Omega -
\kappa/2$ rise to high values near the centre, requiring any linear
bisymmetric modes to have very high pattern speeds, small corotation
radii, and to have very low growth rates (because the inner disc is
not all that responsive).

The outer disc, on the other hand, is highly responsive but has no
cavity-type modes.  We see evidence for weak edge-type modes, which
arise from a steep density gradient \citep{Toom81} at the sharply
truncated outer edge, but they are sufficiently far out and of low
enough frequency to be decoupled from the bar forming process in the
inner disc.

Shot noise from the particles is vigorously amplified, but transient
swing-amplified responses should be damped at the inner Lindblad
resonance (ILR) of the disturbance \citep[][p.~510]{Toom81,BT08}, as
long as the amplitude remains tiny.  Large amplitude waves are not
damped, however, and trap disc particles near the ILR into a bar-like
feature \citep{Sell89}.

\begin{figure}
\includegraphics[width=.57\hsize,angle=270]{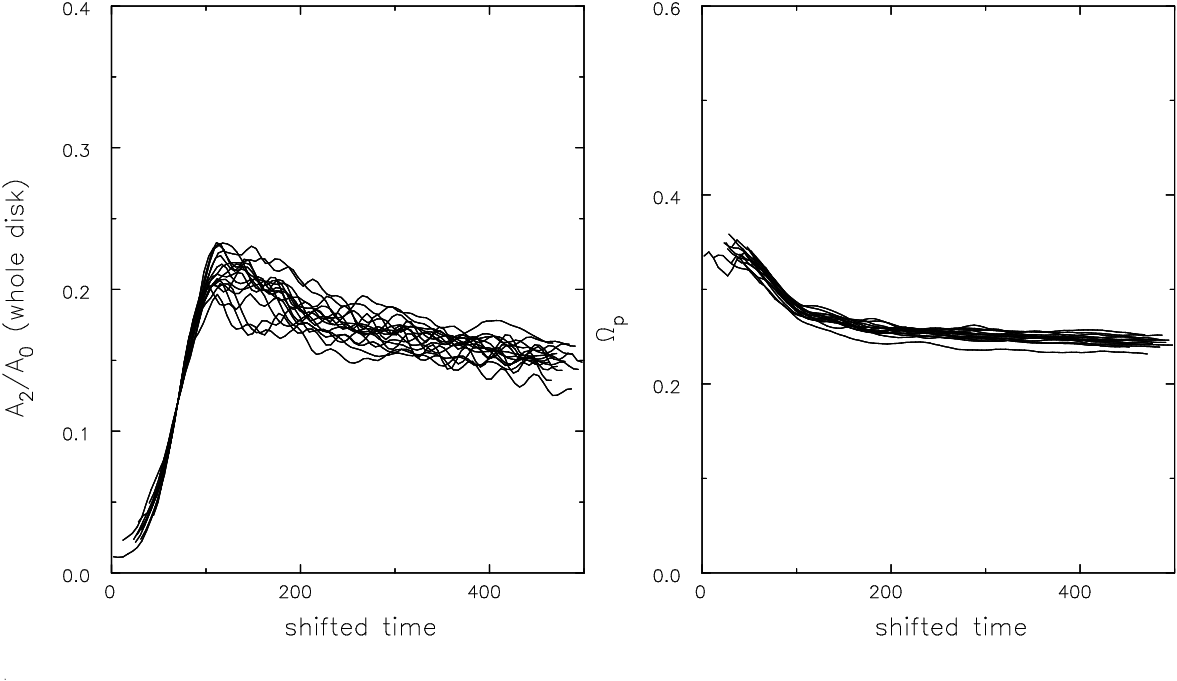}
\includegraphics[width=.57\hsize,angle=270]{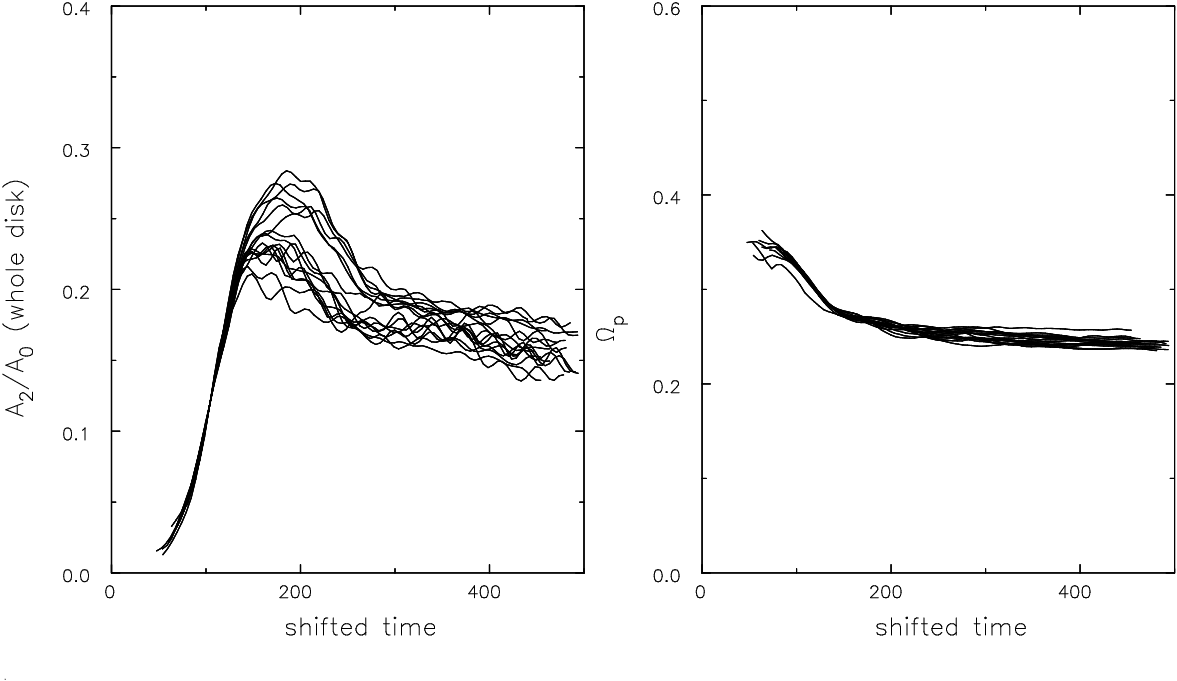}
\includegraphics[width=.57\hsize,angle=270]{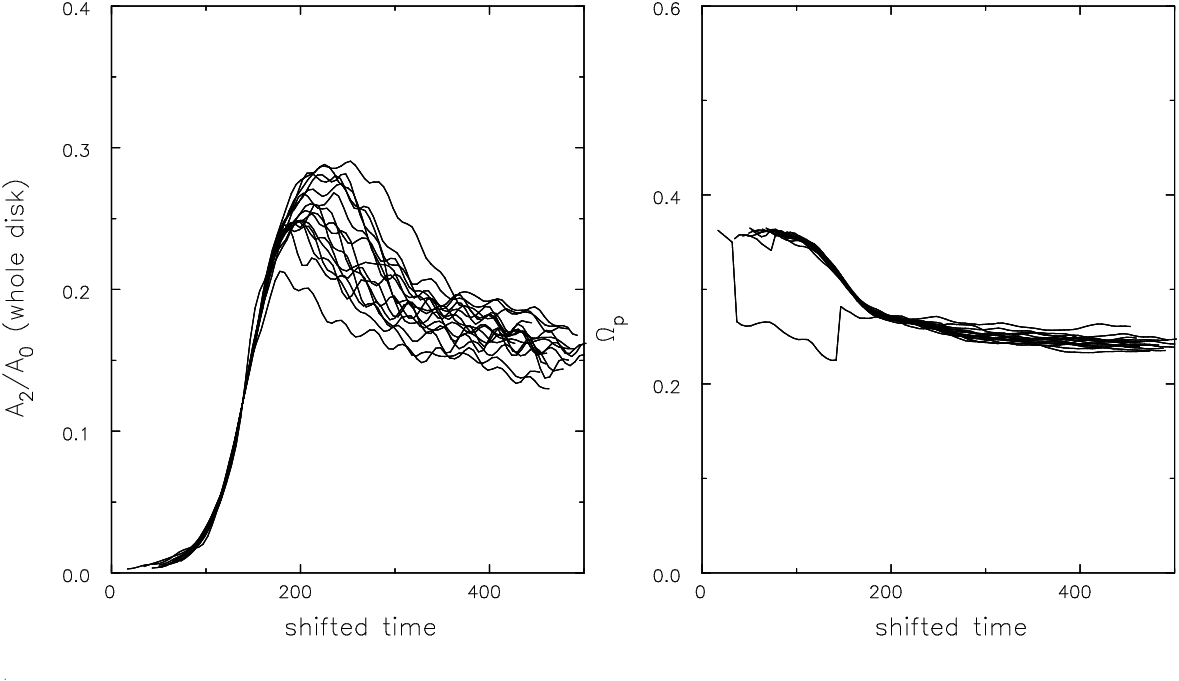}
\includegraphics[width=.57\hsize,angle=270]{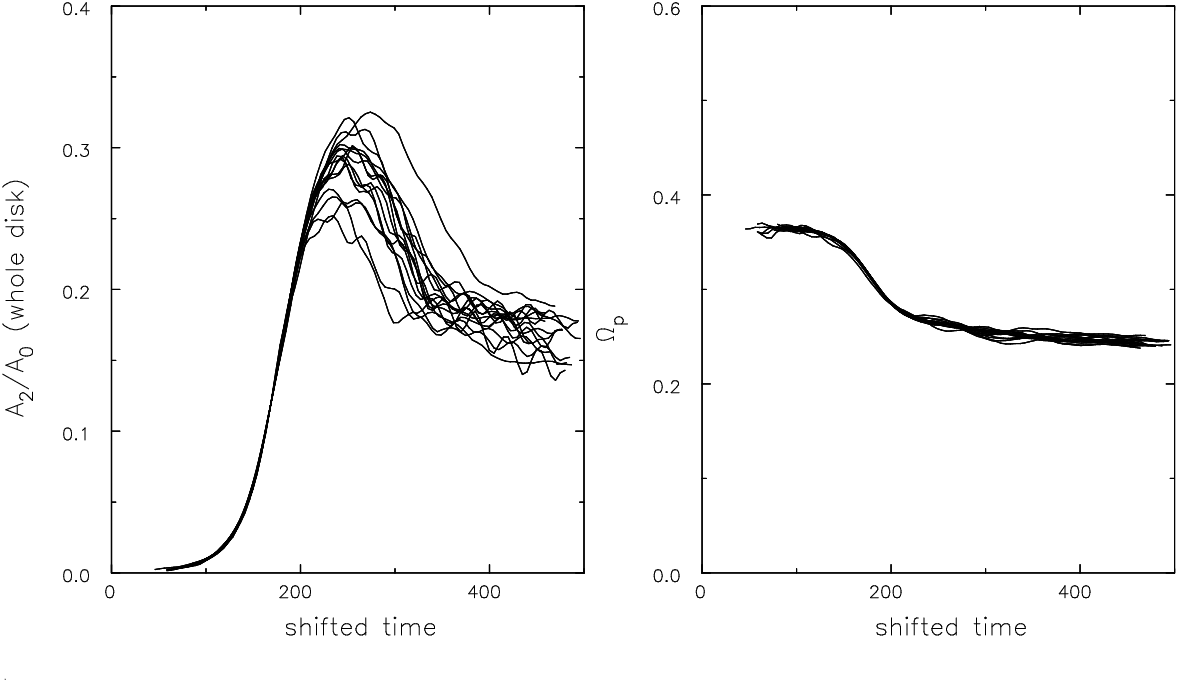}
\caption{Evolution of the bar in four sets of 16 runs with different
random seeds for the disc particle coordinates.  The number of
particles rises by a factor of 10 from row to row, ranging from 50K in
the top row to 50M in the bottom row.}
\label{p2dNvary}
\end{figure}

\begin{figure}
\includegraphics[width=.57\hsize,angle=270]{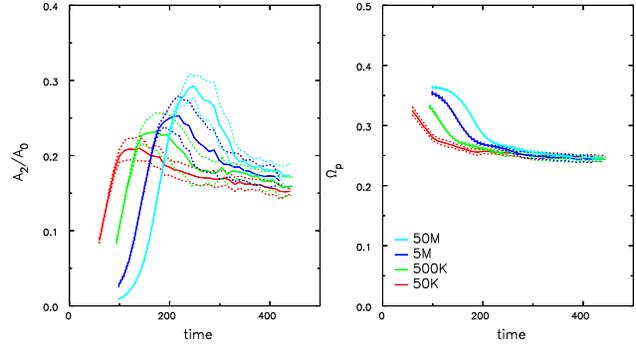}
\caption{Summary plot showing the means and $\pm1\sigma$ scatter of
the runs shown in Fig.~\ref{p2dNvary}.}
\label{multiN}
\end{figure}

Bar formation through amplified noise inevitably leads to a range of
bar properties, but it is fortunate that the range turns out to be
surprisingly narrow.  To illustrate this, we study bar formation in
our standard model in simplified simulations in which the motions of
disc particles are confined to a plane, and the halo particles are
replaced by a rigid mass component that simply provides the extra
central attraction to yield the same rotation curve as shown in
Fig.~\ref{rchern}.  This approach has several advantages: the
calculations are less expensive in computer time, but more importantly
the dynamics is simpler because both bar buckling and halo friction
are eliminated, enabling us to isolate the bar formation process from
these other complicating aspects of the overall evolution.

Fig.~\ref{p2dNvary} shows 4 sets of 16 runs each in which $N$ is
increased by a factor 10 from row to row, from $N=50$K at the top, to
$N=50$M for the bottom row.  The results from each run have been
slightly shifted horizontally so that the amplitude passes through 0.1
at the same time (the mean for the 16 runs) as described above.  The
bar amplitude has a higher peak than in Figs.~\ref{basichyb} \&
\ref{pkdgrav} in part, at least, because we use a different softening
rule in 2D.  The discrepant line in one of the pattern speed panels
shows that the bar cannot always be identified in the early stages,
but eventually it is in all cases.

Fig.~\ref{multiN} shows the evolution of the means and scatter in
the four sets of experiments, and reveals that the main effects of
increasing $N$ are threefold: the formation of the bar is delayed
because of lower seed noise, the mean peak bar amplitude increases and
the scatter in the amplitude evolution {\it rises\/} with increasing
particle number, at least to $N=5\,$M.  The pattern speeds are better
behaved, with scatter decreasing as $N$ rises.

Because these calculations have less freedom, the amplitude variation
is much less than those shown in Fig.~\ref{basichyb}, which have
the same numbers of disc particles as those in the second row of
Fig.~\ref{p2dNvary}.  Nevertheless, the spread in the bar amplitudes
after the initial rise remains quite high.  The pattern speed does not
decline as much because the rigid halo does not cause dynamical
friction.

Since amplified noise is intrinsically stochastic, the dominant
transient responses in different random realizations of the disc must
differ.  The possible frequency range of the dominant pattern is
broad, but not unbounded; the rotation curve and surface density
profile, among other properties, cause the responsiveness of the disc
to vary with radius, and therefore the dominant responses have
corotation radii in the region where the disc is most responsive.
Thus the very first collective responses at low, but fixed, $N$ lead
to initial bars having a range of strengths, \ie\ sizes, with the
larger bars developing more slowly because the clock runs more slowly
farther out in the disc.  (The time delays have been removed from
Fig.~\ref{p2dNvary}.)

The larger the number of particles, the longer it takes for the bar to
form (Fig.~\ref{multiN}).  Initial transient responses occur at
roughly the same rate but, in experiments with larger $N$, the lower
initial amplitudes do not lead to immediate bar formation.  Subsequent
amplification events tend to be of greater amplitude, and to occur
farther out in the disc.  Thus we see that a lower level of shot noise
favours large amplitude responses farther out in the disc that briefly
lead to longer and stronger bars.

The pleasant surprise is that after the initial transient episodes
produce bars of different sizes and angular speeds, we observe
(Fig.~\ref{multiN}) that subsequent evolution causes the range of
bar strengths to narrow.  Also most of the systematic trends with
particle number are erased in the subsequent evolution, and neither
the bar amplitude nor its pattern speed at later times exhibits more
than a mild dependence on $N$.  It is fortunate that a degree of
uniformity of the bar properties emerges after such tumultuously
different evolution.  But it is far from obvious why it should,
especially since the model could have supported bars of wide range of
sizes (\eg\ Fig.~\ref{basichyb}).

The results shown in Fig.~\ref{p2dNvary} are for models with rigid
haloes in which the disc was created using the Jeans equations
(Section~\ref{jeansapp}).  Far from becoming better behaved, the scatter in
the amplitude evolution {\it increases\/} as $N$ rises!  We conducted
a similar set of tests, also with rigid haloes, for which disc
particles were selected deterministically from an approximate \DF.
The evolution of these more carefully set up models resulted in
slightly improved behaviour: the bar formed somewhat more slowly,
peaked at a little lower amplitude for the same value of $N$, and the
scatter no longer varied systematically with $N$.  However, the final
bar amplitude and pattern speeds were within the ranges shown in
Fig.~\ref{p2dNvary}.

Unlike the results for the isochrone presented in Appendix C, the more
careful selection of particles yielded only a slight reduction in the
spread in evolution.  It is likely that this difference in behaviour of
the two discs is due to the difference in bar forming mechanism; the
instability of the isochrone disc is due to strongly unstable linear
global modes, whereas as the bars in our standard model form through
non-linear trapping of swing-amplified particle noise that would be
less affected by the quality of the equilibrium.

Thus far we have discussed only bisymmetric instabilities, but other
low-order instabilities may also be competitive.  In fact, we find
some evidence for lop-sidedness, which we describe in the next
subsection.

\begin{figure}
\includegraphics[width=.57\hsize,angle=270]{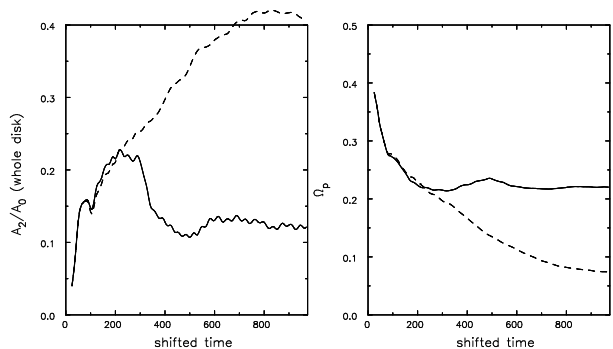}
\caption{Comparison of the time evolution of two runs that differ only
in the imposition of reflection symmetry about the midplane.  The
solid curves are for a model taken from Fig.~\ref{basichyb} in which
vertical forces are unrestricted while the dashed curves show the
evolution of the same initial model when vertical forces from the disc
are constrained to be symmetric about the mid-plane.}
\label{rflctz}
\end{figure}

\subsection{Bending modes}
The bars in most 3D simulations suffer from buckling instabilities
that, when they saturate, thicken the bar in the vertical direction
\citep[\eg][]{CS81,RSJK}.  In many, but not all, cases the evolution
of this bending mode is quite violent and weakens the bar
significantly, while the central density of the bar rises, as reported
by Raha \etal\ \ The radial rearrangement of mass evidently liberates
the energy needed to puff up the bar in the vertical direction.

The time of saturation of the buckling mode depends on a variety of
factors, such as the formation time of the bar, and the initial seed
amplitude of the bending mode, the strength of the bar, \etc\ Several
of these factors will in turn depend on the already stochastic
formation of the bar.  It is hardly surprising therefore, that this
event occurs over a wide range of times and with a wide range of
severity (Fig.~\ref{basichyb}), thereby compounding the overall
level of stochasticity.

The buckling mode can be inhibited by artificially imposing reflection
symmetry about the mid-plane, which causes a substantial change to the
evolution.  Fig.~\ref{rflctz} compares the evolution for one case;
the dashed curves show that when buckling is inhibited, the bar
continues to grow in amplitude, while slowing, for a long period.  On
the other hand, the amplitude drops quite abruptly when the bar
buckles (solid curves) and the subsequent amplitude and pattern speed
hold approximately steady.

\begin{figure}
\includegraphics[width=.57\hsize,angle=270]{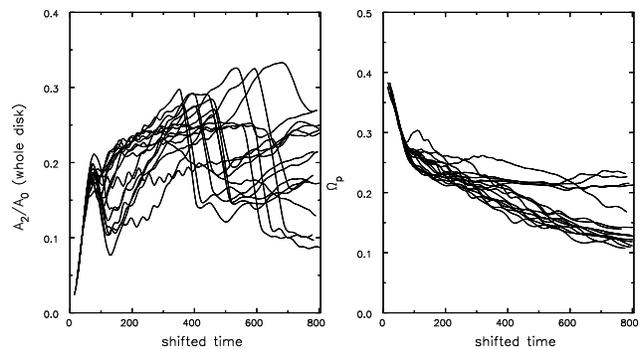}
\caption{Evolution of a set of 16 runs that differ from those shown in
Fig.~\ref{basichyb} only in the suppression of lop-sidedness about the
$z$-axis. }
\label{nomeq1m}
\end{figure}

Not all the bars in the runs illustrated in Fig.~\ref{basichyb}
experience a violent buckling event.  In some cases the bar amplitude
does not decrease after the initial peak, while in others the
amplitude drop is more gradual.  

Fig.~\ref{nomeq1m} shows the effect of suppressing the $m=1$
sectoral harmonic about the $z$-axis for both the disc and halo
particles.  This has the effect of preventing the centres of either
component from leaving the $z$-axis.  (Suppressing the $l=1$ component
of the halo force calculation would nail the centre of that component
to the origin, which would prevent the halo from responding properly
to a buckling mode.)  With lop-sidedness inhibited in this way, all
bars buckle, and all but one do so violently with a large decrease in
amplitude.  This difference in buckling behaviour from that shown for
the same initial models in Fig.~\ref{basichyb} indicates that
buckling is strongly influenced by mild lop-sidedness, which has not
been reported elsewhere, as far as we are aware.  We could not find
any evidence for lop-sided instabilities in the runs shown in
Fig.~\ref{basichyb}, and the distance between the centroids of the
halo and disc particles was $\la 0.002R_d$.  As it seems unlikely that
such small offsets could have such a large effect on the saturation of
the buckling mode, we think it possible that an anti-symmetric mode
competes.  Investigation of this possibility here would be too great a
digression.

Despite the violence of most buckling events, most bars in these
restricted simulations continue to slow after the buckling event and
amplitude growth resumes.  The four exceptions are bars that remained
strong right after their formation and did not slow much either before
or after the buckling event.

Results reported in Appendix B show that the buckling behaviour is also
somewhat sensitive to particle softening.

\citet{KVCQ} report that the violence of the buckling event also
depends on the initial thickness of the disc.  This is as expected,
since \citet{MS94} showed that buckling is a consequence of a
collective instability that arises in systems in which the velocity
distribution becomes too anisotropic, and thickening the disc reduces
the flattening of the velocity ellipsoid.  However, in a separate test
with a set of runs with twice the disc thickness (not shown), we still
find a similar degree of scatter in the late evolution.

\begin{figure}
\begin{center}
\includegraphics[width=.57\hsize,angle=270]{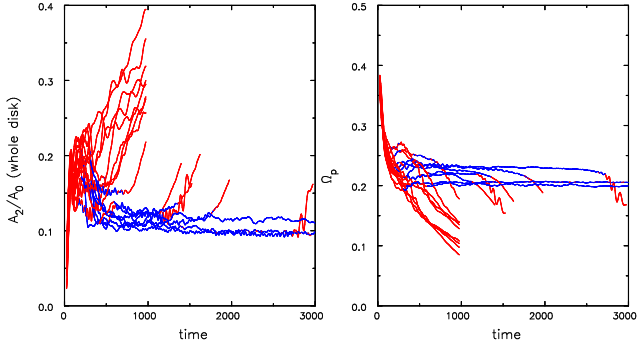}
\end{center}
\caption{The results shown in Fig.~\ref{basichyb}, but with the curves
colored blue when the torque on the halo is low and red otherwise.
The calculations were continued for models that had not slowed by
$t=1000$ and were stopped either at $t=3000$ or soon after friction
kicked in, which happened in all but two cases.}
\label{contd}
\end{figure}

\subsection{Incidence of Dynamical Friction}
\label{dynfr}
Fig.~\ref{contd} shows that the divergent late-time evolution of the
runs shown in Fig.~\ref{basichyb} is due to differences in the
incidence of dynamical friction.  The lines are coloured blue when the
torque acting on the halo $dL_z/dt < 5 \times 10^{-5}GM^2/R_d$, and
are red otherwise.

The absence of bar friction may have a variety of causes: (a) low halo
density, (b) a weak bar, and (c) metastability caused by local adverse
gradients in the density of halo particles as function of angular
momentum \citep{SD06}.  The halo density is just about the same in all
cases, but the bar strength varies widely and it is clear that the
weaker bars experience little friction.

\begin{figure*}
\begin{center}
\includegraphics[width=.53\hsize,angle=270]{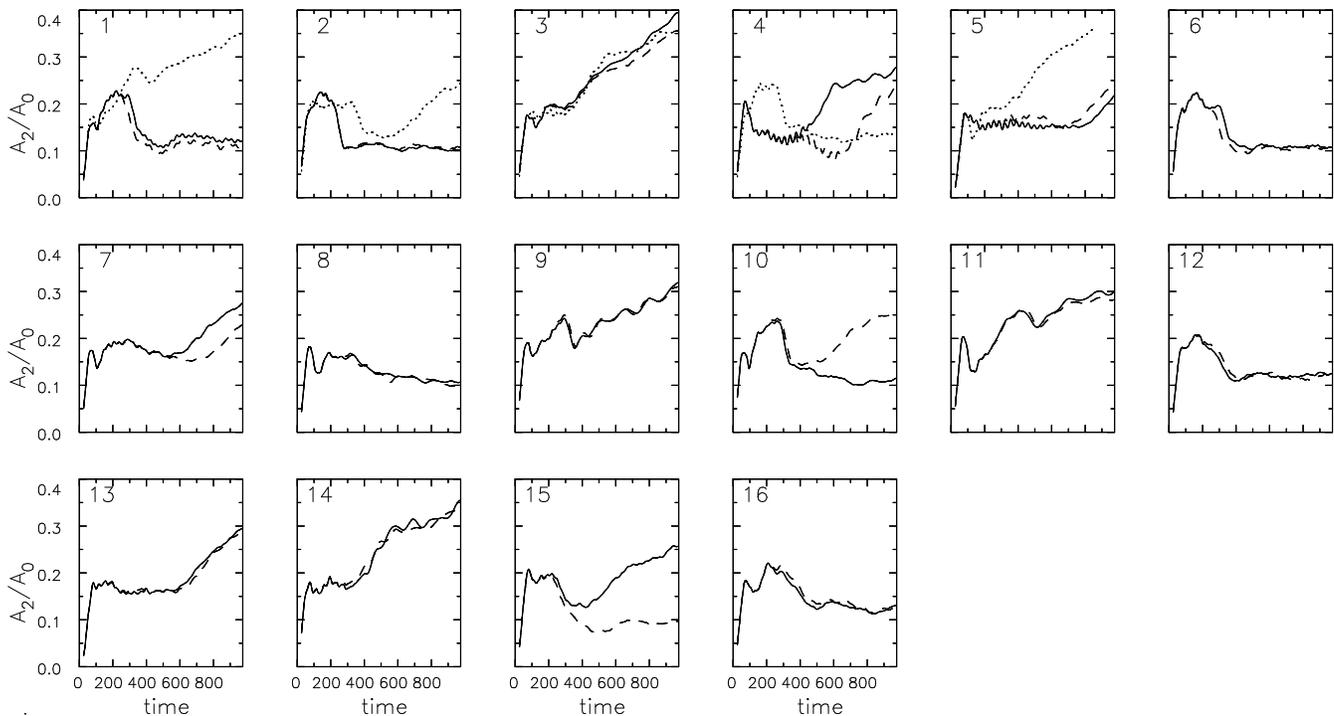}
\end{center}
\caption{Comparison of the amplitude evolution of the models shown in
Fig.~\ref{basichyb} (solid lines) with the same sets of particles
processed in reverse order (dashed lines).  The evolution of these
two sets of identical runs is measurably different in all cases, and
qualitatively different in some, especially cases 10 \& 15.  The dotted
lines in the first 5 panels show the evoltions using \PK\ for the same
files of initial particles.}
\label{cmprsn}
\end{figure*}

The third possibility is indicated by the evidence in
Fig.~\ref{contd}, since friction eventually resumes, sometimes after
a very long period during which the bar amplitude does not increase;
the metastable state does not last indefinitely.  We argue
\citep{SD06} that the metastable state has a finite lifetime because
weak friction at minor resonances gradually slows the bar until the
more important resonances move out of the region of adverse gradients,
allowing strong friction to resume.

Metastability could be caused by the buckling event, since bars that
are weakened substantially by a buckling event, such as the case
picked out in Fig.~\ref{rflctz}, generally do not experience much
friction at late times, and their amplitudes stay low.  The upward
rise in the bar pattern speed at the time of buckling is shown clearly
by the solid curve in Fig.~\ref{rflctz}, which we \citep{SD06} found
to be a likely cause of metastability.  It is reasonable that the
concentration of mass to the centre as the bar buckles should cause an
upward fluctuation in the bar pattern speed (because the orbit periods
must vary inversely as the square root of the mean interior density).
However, buckling does not always lead to a cessation of friction; for
example, many of the bars in the 16 runs with a more dense halo
(Fig.~\ref{newhalo}) clearly buckled, but friction continued in all
cases.

\subsection{True chaos}
Here we show that Miller's (1964) instability can lead to macroscopic
differences in discs.  Where initial evolution is largely determined
by swing-amplification of the spectrum of particle noise laid down by
the random coordinates of particles, models that differ by tiny
amounts quickly diverge because the subsequent spiral events depend on
the details of evolution of previous events.  This phenomenon causes
the micro-chaos in $N$-body systems to lead to macroscopic differences
in discs.

Fig.~\ref{cmprsn} compares the amplitude evolution of each case
shown in Fig.~\ref{basichyb} (solid lines) with another run of the
same case with the order of the particles reversed (dashed).  Thus the
initial phase space coordinates of all particles were identical and
are evolved with the same code on identical processors.  Each pair of
simulations differ only in the order in which arithmetic operations
are performed, which changes the initial accelerations at the round
off error level only, yet the amplitudes at late times generally
differ visibly, and in some cases, \eg\ 10 \& 15, the evolution
differs qualitatively.

So far, every calculation with grid codes that we have reported here
was conducted using single precision arithmetic for most operations.
We have checked that increased precision has no effect on the range
of behaviour shown in Fig.~\ref{basichyb}, and results differ only
slightly, as we now show for one case.

\begin{figure}
\includegraphics[width=.57\hsize,angle=270]{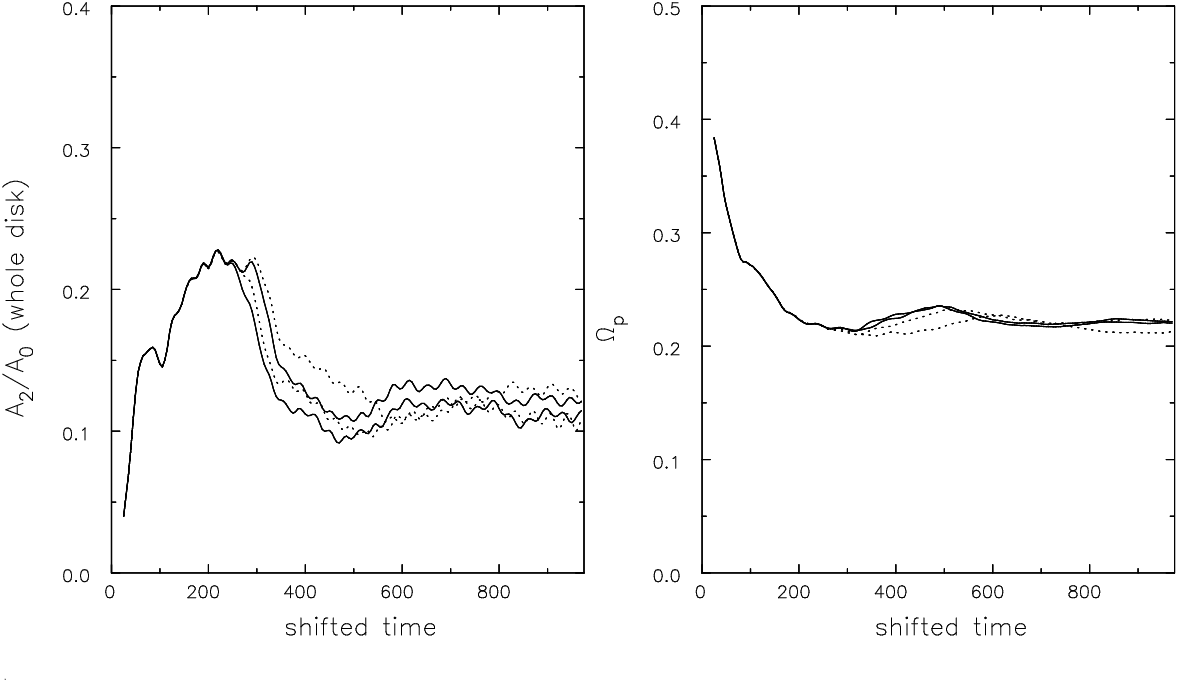}
\centerline{\includegraphics[width=.9\hsize]{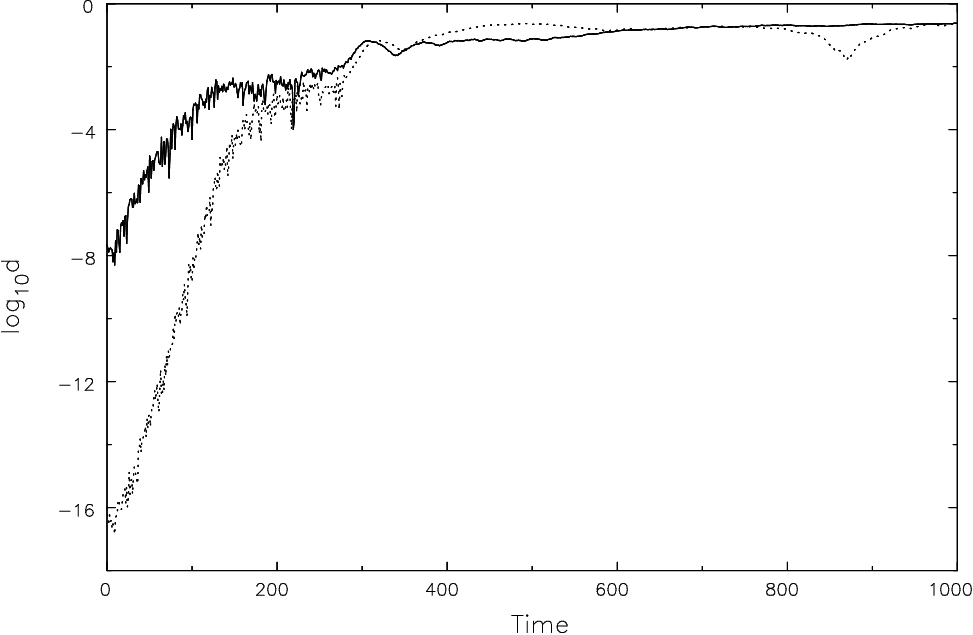}}
\caption{The upper panels compare the evolution of four cases that
started from the identical file of particle coordinates, with all
numerical parameters held fixed, except that solid lines are for
calculations in single precision, dotted lines are for the identical
calculations in double precision.  As for Fig.~\ref{cmprsn}, the order
of the particles was reversed in one of each pair.  The lower panel
shows the time evolution of the quantity $d$ defined in
eq.~(\ref{diff}) for both pairs of runs.}
\label{lyapunov}
\end{figure}

Fig.~\ref{lyapunov} shows that the system remains chaotic when we
repeat the calculations using double precision arithmetic (dotted
lines).  The higher precision calculations begin to diverge visibly at
about the same times as in the single precision cases, and the
subsequent differences are comparable.  In order to monitor the
divergence in these cases, we compute the value over time of the
difference
\begin{equation}
d = \left[ \Re(A_{2,a} - A_{2,b})^2 + \Im(A_{2,a} - A_{2,b})^2 \right]^{1/2}
\label{diff}
\end{equation}
between the bar coefficients (eq.~\ref{fcoeff}) in these pairs of
experiments ($a \; \& \; b$) in which the order of the particles was
reversed.  The solid (dotted) line in the lower panel of
Fig.~\ref{lyapunov} shows the result for the single (double)
precision pair.  By $t \sim 300$ the models differ quite visibly in
the amplitude and phase of the bar, which accounts for the fact that
$d$ asymptotes to a lasting value where the phases of the two bars
differ.

The difference, $d$, in double precision grows quasi-exponentially
over time at first, which is symptomatic of chaos, with a Lyapunov
($e$-folding) time of $\sim 4.75$ dynamical times, \ie, less than 25\%
of the orbit period ($\sim 20$ dynamical times) at $R = 2.5R_d$.
Using this estimate of the Lyapunov time, the difference in the double
precision case should equal the initial difference in the single
precision case after $\approx 93$ dynamical times, and the early
evolution of $d$ in the lower precision case is roughly similar to
that in the double precision case with a time offset of this
magnitude.  Even though there is a much smaller initial difference
between the two double-precision models, the seed amplitude of the
instabilities is set by the shot noise, which is the same in all 4
runs.  Thus the non-axisymmetric structures are almost fully developed
in the double precision models by the time the dotted curve reaches
the level of the start of the solid line; therefore one cannot expect
the curves to overlay perfectly.

It is curious that the difference in the double precision case
``catches up'' with that in the single precision case.  The shoulder
in $\log_{10}d$ that appears in both precisions at about $t=300$ seems
to be responsible for this convergence, which occurs both at such a
large value of $d$ as to be well past where exponential divergence
could be expected to hold, and at a time when the bar in all four runs
is fully developed.

\begin{figure}
\includegraphics[width=\hsize,angle=0]{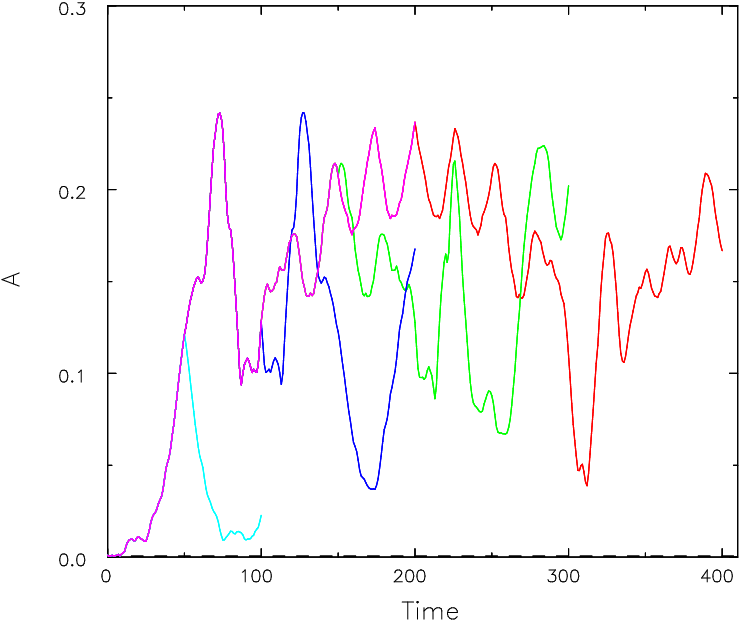}
\caption{The magenta line shows the unsmoothed bar amplitude evolution
of one model run to $t=200$.  The other lines show the continued
evolution of the same model with the velocities of all the particles
reversed at $t=50$ (cyan), $t=100$ (blue), $t=150$ (green), and
$t=200$ (red).  In all four cases, the evolution immediately after the
reversal faithfully retraces the forward evolution for a period less
than 100 time units.  After this time the evolution departs noticeably
from a reflection of the line about the reversed moment.}
\label{revers}
\end{figure}

A perfect collisionless particle system should be exactly time
reversible; that is, if the velocities of all the particles were
reversed at some instant, the system should retrace its evolution.
Fig.~\ref{revers} shows that reversed simulations do retrace their
evolution for a short while, between 60 and 80 dynamical times, after
which the evolution of the reversed model visibly departs from the
corresponding reflection of the forward evolution.  This period of
successful reversibility is consistent with our Lyapunov divergence
estimate: 15 Lyapunov times ($=71.25$ dynamical times) corresponds to
a divergence of $\sim 10^{6.5}$, which is sufficient to alter almost
every significant digit in these single precision calculations and
lead to reversed evolution that becomes largely independent of that in
the forward direction.  Further analysis of these simulations revealed
that the first signs of irreversibility appeared as differences in the
leading spiral Fourier components, suggesting that vigorous
swing-amplification of particle noise is primarily responsible for the
short Lyapunov time.

We conclude from these tests that the $N$-body system we are trying to
simulate is indeed chaotic.  Further, the effects of chaos are not
significantly worsened by round-off error in single precision; we
have also verified that the full divergence of the results in
Fig.~\ref{basichyb} persists in double precision.  In fact, the
first author has frequently checked, and always confirmed, that no
advantage results from use of higher precision arithmetic when
computing the evolution of collisionless $N$-body systems.  This
conclusion is in sharp contrast with the requirements for collisional
systems \citep[\eg][]{Aars08}.

In none of the simulations with grid codes reported in this paper did
we distribute the computation over multiple parallel processors, even
though the code has been well optimized for parallel use.  We adopted
this strategy in order to avoid the additional randomness that is
inevitable when results from multiple processors are combined in an
unpredictable order.

The dotted curves in the first five panels of Fig.~\ref{cmprsn} show
the result using the tree code \PK\ for the same initial coordinates
in each case, which are reproduced from Fig.~\ref{pkdgrav}.
Although the ranges and distributions of measured bar properties shown
in Figs.~\ref{basichyb} \& \ref{pkdgrav} are similar, the results do
not compare in detail, as noted above.  Results from the two different
codes diverge strongly in all but one case, reinforcing the conclusion
of intrinsic stochasticity.  Which of the two possible evolutionary
paths is taken in the evolution is affected no more, and no less, by
code differences than by choices of the random seed.

\section{Discussion}
\label{discussion}
\subsection{Is there a right result?}
One of the most troubling aspects of the diverging evolution in
Figs.~\ref{basichyb} \& \ref{pkdgrav} is that one cannot decide
which of the two patterns of behaviour is ``correct,'' or indeed
whether there could be a unique evolutionary path with a perfect code
and infinite numbers of particles.

Since these models have high density centres (Fig.~\ref{rchern}),
linear stability analysis would most likely reveal that all global
modes, with the possible exception of edge modes \citep{Toom81}, have
very low growth rates, and therefore the disc ought to be stable and
not form a bar.  If this is indeed what linear theory would predict,
then the ``right result'' with a perfect code and infinite numbers of
particles would be a stable model that does not form a bar.  This
outcome never occurred in the $> 400$ simulations we report here,
even in cases with one hundred times our standard number of disc
particles (Fig.~\ref{p2dNvary}).

The level of shot noise in a simulation with $\ga 1$~million particles
is clearly $\sim 100$ times higher than would be present in a real
galaxy if the $\sim 10^{10}$ stars were randomly distributed.  But the
mass in real galaxy discs is clumpier because of the existence of star
clusters and giant gas clouds, which raises the amplitude of random
potential fluctuations -- although the density fluctuation spectrum
may not be the same as that of shot noise in the simulations.
Nevertheless, it seems most unlikely that a real galaxy closely
resembling the model used in our simulations could avoid being barred.

\subsection{Dynamical Friction}
The greatest source of divergence is the bimodal nature of dynamical
friction, which is avoided for a long time in some cases, but kicks in
immediately in others, causing the bar to slow and increase in
strength by a substantial factor.  It is likely that friction is
avoided because the needed gradient in the halo \DF\ as a function of
angular momentum has been flattened by the earlier evolution of the
model, as reported by \citet{SD06}.  The fact that this happens here
more frequently than we found with the model created by \citet{VK03}
may have two causes: their model had both a less dominant disc and an
initial halo with significant departures from equilibrium.

In Section~\ref{pselect}, we reported a weak trend towards a larger
fraction of non-slowing bars as we took greater care over the initial
selection of particles; further, the largest fraction (10/16) occurs
in the test with four times the number of halo particles reported in
Appendix B.  This weak trend suggests that the metastable state is
reached more readily as the quality of the simulation is improved.

However, \citet{SD06} found that the metastable state, in which the
bar did not slow, was not indefinite and friction eventually resumed,
as we also find here (Fig.~\ref{contd}).  Furthermore, they found
the metastable state to be fragile, and friction would resume soon
after a tiny perturbation, such as the distant passage of a small
satellite galaxy.  Thus, even though the metastable state is reached
more frequently in higher quality calculations, it is unlikely it
could be sustained in real galaxies.  We conclude therefore that the
strongly braked and growing bar is the most ``realistic'' outcome from
these simulations.

\subsection{Introducing a seed disturbance}
\citet{HBWK} attempted to make the outcome more predictable by seeding
the bar instability by an externally applied transient squeeze.  We
argue here that this approach is not the panacea it may seem.

In the case of discs having well-defined global instabilities, noisy
starts already seed the dominant unstable modes at high amplitude
\citep[Section~\ref{multimodes};][]{Sell83}.  If a seed disturbance is to
prevail, it must be imposed at such high amplitude as to be
practically non-linear at the outset.  Furthermore, the objective must
be to favour the dominant mode over the others, which cannot be
achieved by a simple perturbation.  Instead, one must impose both the
detailed radial shape and perturbed velocities of the mode, which are
generally not known.  A more generic disturbance, such as a
``squeeze'' will simply raise the amplitude of all the modes and
transients, giving {\it less\/} time for the dominant mode to outgrow
the others.  Quiet starts \citep[Section~\ref{qstart};][]{Sell83,SA86},
however, have the effect of reducing the initial amplitudes of all
non-axisymmetric disturbances to such an extent that there is ample
time for the most rapidly growing mode to prevail.  Thus the outcome
of a quiet start experiment is tolerably reproducible without the need
to apply an additional seed (Fig.~\ref{isoc8dfdetq}).

The situation is far more difficult in the case, as in the present
study, where the disc has no prevailing global instabilities, since
the evolution of a simulation is dominated by swing-amplified shot
noise.  Quiet starts are all but useless in these circumstances also,
since they break up rapidly as the tiny seed noise is swing amplified,
with similar outcomes, only slightly delayed, to those from noisy
starts.  Cranking up the particle number does not reduce variations in
the bar amplitude at later times (they actually increased in
Fig.~\ref{p2dNvary}), but does delay bar formation.  Because of
this, perhaps a suitable seed disturbance in a very large $N$ disc may
prevail over the amplified shot noise and lead to a more reproducible
outcome.  We have not explored this idea here and leave it for a
future study.

\section{Conclusions}
We have shown that simulations over a fixed evolutionary period of
a simple disc-halo galaxy model can vary widely between cases that
differ only in the random seed used to generate the particles, even
though they are drawn from identical distributions.
Fig.~\ref{basichyb} shows that the late-time amplitude of the bar
can differ by a factor of three or more while the stronger bars may
have half the pattern speed of the weaker ones.  Fig.~\ref{contd}
shows that the largest differences are only temporary, however.  We
have deliberately focused our study on a case which displays this
extreme bad behaviour.  Stochastic variations are inevitable, but
evolution is generally less divergent; \eg, when the halo has both
higher and lower density (\eg\ Fig.~\ref{threehalos}).

We have shown that the divergent outcomes do not result from a
numerical artefact, since they are independent of numerical parameters
(Appendix B).  Also, similar behaviour occurs with a code of a totally
different type (\PK, see Fig.~\ref{pkdgrav}).  Instead, this extreme
stochasticity results from a number of physical causes that we have
identified and illustrated.  The most important for our model are:
swing-amplified particle noise, the variations in the incidence and
severity of buckling, and the incidence of dynamical friction.  We
have separately shown (Fig.~\ref{isoc8dfdetn}) that other disc
models having a well-defined spectrum of global modes can have a range
of outcomes because of the coexistence of competing instabilities.

The calculations in Fig.~\ref{basichyb} are of models that were set
up with considerable care so as to be as close as possible to
equilibrium.  An additional level of unpredictability can result from
less careful set-up procedures, as illustrated in Appendix C.

We have been aware for many years that simulations including disc
components can be reproduced exactly only if the arithmetic operations
are performed in the same order to the same precision, and that
differences at the round-off error level can lead to visibly different
evolution.  However, we have been surprised by the strongly divergent
behaviour of the particular model studied here.  The pairs of
divergent results in Fig.~\ref{cmprsn} are the stellar dynamical
equivalents of the possible macroscopic atmospheric consequences of
Lorenz's butterfly flapping its wings.  Because the system is chaotic,
improved precision arithmetic is of no help in reducing the scatter in
the outcomes.

The divergence in different realizations of our standard case arises
from a temporary delay in the incidence of dynamical friction, which
is determined by minor details of the early evolution.  Strong
friction causes the bar to both slow and grow; in some cases this
occurs right after bar formation, but in others the bar rotates
steadily at an almost constant amplitude for a protracted period.
Friction is avoided when the earlier evolution causes an inflexion in
the angular momentum density gradient of the halo.  We \citep{SD06}
previously described this as a metastable state because it did not
last indefinitely even when the evolution was unperturbed, and we also
showed that mild perturbations could cause friction to resume.  We
find that the fraction of initially non-slowing bars increases as
greater care is taken over the initial set up because the smaller
fluctuations in such models are less likely to nudge the model out of
the metastable state.

We argue in Section~\ref{discussion} that the most realistic outcome of
these experiments is the slowing and growing bar, despite the fact
that we find the delayed friction result increasingly often as we
improve the quality of the initial set-up and of the simulation.
Since most real galaxies are likely to be subjected to frequent mild
perturbations, we conclude that slowing and growing bars are in
fact the more realistic outcome.

Since the possible evolution of the simulation is not unique, multiple
experiments of essentially the same model are needed in order to
demonstrate that the behaviour is robust.  Furthermore, the failure of
an experiment by one group to reproduce the results of a similar
experiment by another may not be the result of errors or artefacts in
either or both codes, but rather a reflection of a fundamental
stochasticity of the system under study.

\citet{KVCQ} report a similar, but less extensive, comparison
between two tree codes and an adaptive mesh method, and conclude that
all the codes produce ``nearly the same'' results in simulations
performed with sufficient numerical care.  However, inspection of the
comparatively short evolution shown in their Fig.~8 reveals slowly
diverging outcomes, even between two simulations run with tree codes.
They also report (their Fig.~1) a strongly divergent result when the
time step was varied; the sharp decrease in bar strength in this one
case was clearly a consequence of a more violent buckling event than
in their comparison cases.  Such a difference could have easily arisen
from stochastic variations of the kind discussed here, and the
conclusion that the shorter time step is required no longer follows.
We show here (Appendix B), as do \citet{DBS9}, that results are robust
to wide variations in time step.  Clearly when stochasticity can lead
to sharply divergent results, parameter tests that throw up surprises
are conclusive only after ensembles of particle realizations have been
simulated.  This must also be a requirement for meaningful comparisons
between codes or workers.

Since the principal sources of stochasticity are connected to disc
dynamics, they are unrelated to the halo particle number question
raised by \citet{WK07}.  Not only has \citet{Sell08} already shown
that friction can be captured adequately with moderate particle
numbers, but we have found here that the expected bar friction arises
more readily in haloes with fewer or equal mass halo particles, or in
haloes that are not set up with great care -- which is not the
expected behaviour were particle scattering dominant.  Instead, small
departures from equilibrium can upset the delicate metastable state in
which bars can rotate without friction \citep{SD06}.

It should be noted that bars that slow through dynamical friction also
grow in length, as reported earlier by \citet{Atha02}.  Nevertheless,
for these models the ratio of corotation radius to bar semi-axis
${\cal R} > 1.4$, as expected for a moderate-density halo \citep{DS00}.
Those bars that avoid friction for a long period, however, have ${\cal
R} < 1.4$, as also found by \citet{VK03}, but this metastable state is
fragile and unlikely to arise in real galaxies \citep{SD06}.

Since all $N$-body simulations are intrinsically chaotic, they can be
reproduced exactly only if the same arithmetic operations are
performed in the same order with the same precision, as noted in the
introduction, and borne out in Fig.~\ref{cmprsn}.  These
requirements dictate the use of the same code, compiler, operating
system, and hardware.  Further, if the calculation is stopped and then
resumed, it is important to save sufficient information so that the
acceleration used to advance each particle at the next step is
identical, to machine precision, to that it would have been had the
calculation not been interrupted.  This can be arranged without too
much difficulty, if the calculation is run on a single processor.
However, simulations that distribute work over parallel processors in
computer clusters would be exactly reproducible only if care is taken
to ensure that the work is distributed and the results are combined in
a fully predictable manner.

Provided the divergence is slight, exact reproducibility is of little
scientific interest, although such a capability is useful to the
practitioner.  But when, as described here, the model under test can
have strongly divergent behaviour that arises from differences that
begin at the round off level with the same code on the same machine,
comparison of results between different codes and on different
platforms becomes much less likely to produce agreement, even when the
simulations share the same file of initial coordinates.  It is ironic
that the model used here was in fact that selected as a test case for
code comparison; fortunately, the authors discovered its unsuitability
in time!

\section*{Acknowledgments}
We thank Scott Tremaine, Tom Quinn, and the referee, Martin Weinberg,
for helpful comments on the manuscript and Juntai Shen for
discussions.  This work was supported by grants to JAS from the NSF
(AST-0507323) and from NASA (NNG05GC29G) and by a Livesey Grant from
the University of Central Lancashire to VPD.  The \PK\ simulations
were performed at the Arctic Region Supercomputing Center (ARSC).

\appendix
\section{Codes and softening rules}
\subsection{Force Determination Methods}
The accelerations to be applied to particles in an $N$-body simulation
can be determined in many different ways that fall into two broad
classes.  Direct pair wise summation, usually with a tree algorithm to
improve efficiency, and methods that solve for the gravitational field
over a volume.  Three common methods in the latter category are: (1)
solving a finite difference approximation to the Poisson equation on a
grid, (2) convolution between the source distribution and a Green
function on a grid, and (3) expansion of the field in multipoles, with
either a basis set to represent the radial part or a grid on which the
contributions of interior and exterior masses are tabulated.  Grid and
field methods are far more efficient than tree codes, albeit at the
cost of ease of use and versatility.

All grid methods assign masses to a spatial raster of points and
tabulate the field at the same points.  Sensible interpolation schemes
to treat particles between grid points lead to forces between
particles that decrease smoothly at separations below one grid space,
reaching zero for coincident particles.

Finite difference methods solve an approximation to the Poisson
equation directly, yielding a potential arising from the mass
distribution.  Acceleration components, which have to be estimated
from a finite difference approximation to the gradient operator, lead
to forces that approximate the full Newtonian value at distances of
greater than a few mesh spaces, but which are significantly weaker at
short range \citep[\eg\ appendix of][]{SM94}.

Convolution methods, on the other hand, can be used to compute the
acceleration components directly, without the need to difference a
potential.  The Green function is the force field of a unit mass,
which requires a separate convolution for each coordinate direction.
However, the force law needs to be softened at short range both to
prevent acceleration components from varying so steeply across a grid
cell that simple interpolation rules become inadequate and also to
limit the maximum possible acceleration, particularly where grid cells
become very small near the centres of polar grids.

\subsection{Softening Rules}
Since any arbitrary softening rule can be adopted in convolution
methods, physical considerations can be used to select the optimum
rule for a particular application.  The Plummer softening rule for a
unit mass uses the density profile and potential
\begin{equation}
\rho(x) = {3 \over 4\pi\epsilon^3} (1 + x^2)^{-5/2}; \qquad \phi(x)=
-{G \over \epsilon}(1+x^2)^{-1/2},
\label{Plummer}
\end{equation}
where $x = r/\epsilon$, with $\epsilon$ denoted the softening length.
This rule is optimal when particles are confined to a plane, because
it yields in-plane accelerations that would result if the razor-thin
mass distribution were displaced vertically by the softening length.
The forces can be thought of as approximating those from a disc of
finite thickness since softening affects the dispersion relation for
spiral waves \citep[\eg][]{Vand70,Eric75,Rome92} in much the same way
as does finite thickness.  We therefore employ this rule when particle
motion is confined to a plane.

The disadvantage of the Plummer softening rule in 3D simulations is that it
weakens forces on all scales and other rules that avoid this
shortcoming have become popular.  The precise short-range behaviour is
of little importance for relaxation, since inverse square-law forces
imply scattering is dominated by the cumulative effect of long-range
encounters (\eg\ BT08, p.~36).  For our 3D simulations we adopt the
somewhat clumsy cubic spline density kernel used in the original
version of the tree code {\sc PKDGRAV} \citep{Stad01}, which has the
form
\begin{equation}
\rho(x) = {1 \over 4\pi\epsilon^3} \cases{ 4 - 6 x^2 + 3 x^3 & $0 \leq x
\leq 1$ \cr (2 - x)^3 & $1 \leq x \leq 2$ \cr 0 & otherwise, \cr}
\label{monlat}
\end{equation}
as suggested by \citet{ML85}.\footnote{Their expression omits the
square of $x$ in the first line, but this typo is corrected in
\citet{Mona92}.}  The density has continuous second derivatives, while
the potential is given by a messy expression at short range but is, of
course, that of a unit point mass when $x > 2$.

\subsection{Codes}
For fully 3D simulations, we use the hybrid grid method described
elsewhere \citep{Sell03}.  It solves for the field by convolution on a
3D cylindrical polar grid for the disc particles, with the softening
rule (eq.~\ref{monlat}), while the accelerations of the halo
particles are computed using method (3) of Section A1 on a spherical grid.

We also report a number of results using both polar and Cartesian 2D
grids, where we use the Plummer softening law.

In most experiments, we shift the centre of both grids to a new
location every 16 time steps.  The new centre is location of the
particle centroid \citep{McGl84}.  The estimate of the change in this
location is determined by Newton-Raphson iteration, which is repeated
until the shift at each iteration is less than $10^{-3}R_d$.  This
process is unnecessary when any lop-sidedness in the mass distribution
does not contribute to the accelerations and when Cartesian grids or
tree codes are used.

In addition, we have used the tree code, \PK\ \citet{Stad01}, which
adopts the softening kernel K$_1$ recommended by \citet{Dehn01}.  We
have also conducted a few tests with the numerical parameters of time
step, opening angle, \etc, and found results with this code also
are independent of these choices to a similar level of tolerance.

\subsection{Time steps}
When using grid methods, we adopt a 5-zone time stepping scheme in
which the more slowly moving outer particles have time steps that
increase by a factor 2 from zone to zone.  All particles experience
forces from all others at every step, but forces from particles in
outer zones are interpolated to intermediate times \citep{Sell85}.

\subsection{Measurements of $A$ and $\Omega_p$}
We need to make quantitative comparisons of the bar evolution in
simulations as codes, numerical parameters, or random seeds are
varied.  In particular, we compare the evolution of the overall
amplitude and phase of a bar-like distortion in the disc.  We measure
this quantity by computing
\begin{equation}
A_2(t) = {1 \over N_{d}} \sum_{j=1}^{N_d} e^{2i\theta_j}.
\label{fcoeff}
\end{equation}
where $N_d$ is the total number of disc particles and $\theta_j(t)$ is
the azimuth of the $j$th disc particle at time $t$, reckoned from a
fixed direction through the centre defined as the particle centroid at
that time.  Since the quantity $A$ is complex, the bar amplitude is
$|A_2| = [\Re(A_2)^2 + \Im(A_2)^2]^{1/2}$ and phase $2\theta_b =
\hbox{arctan}[\Im(A_2)/\Re(A_2)]$, with the factor 2 appearing in
order to yield a phase that increases by $180^\circ$ as the
bi-symmetric pattern makes half a rotation.  We measure $A_2$ at
frequent intervals, generally every 0.1 dynamical times.  The pattern
speed of the bar is clearly the time derivative of the phase

We make a smoothed estimate of the amplitude and pattern speed by
fitting a steadily rotating wave to the complex $A_2$ values over a
short time interval, and sliding the time interval forward to follow
the evolution of both quantities.  Our plots of amplitude and pattern
speed are of the smoothed quantities.

\section{Tests of numerical parameters}
As always, we check the extent to which the behaviour depends upon all
numerical parameters.  We have been particularly thorough in the case
of our standard model where our results are so surprising.  Since
simulations with a rigid halo and the disc particles confined to a
plane already show large variations (Fig.~\ref{p2dNvary}), we begin
by presenting checks of these inexpensive simulations.

\begin{figure}
\includegraphics[width=.57\hsize,angle=270]{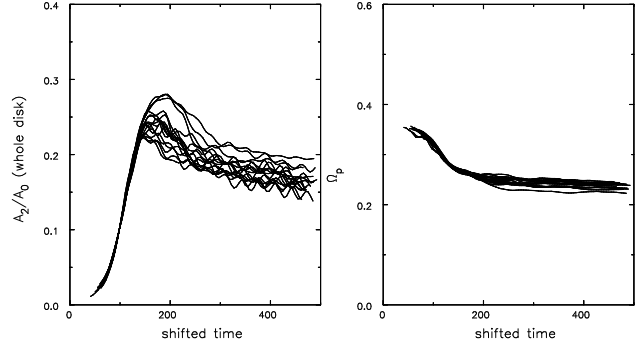}
\caption{Evolution of the bar in 16 runs with different random seeds
for the disc particle coordinates.  These simulations use a 2D
Cartesian grid: numerical parameters are as given in
Table~\ref{pars2d} except $N_x \times N_y = 256^2$, there is a single
time step zone and the grid is not recentred.}
\label{c2d500K}
\end{figure}

\subsection{Grid Geometry}
We have run these calculations on both a 2D polar grid and a 2D
Cartesian grid in order to convince ourselves that our results were
not being affected by our choice of grid geometry.  The result for the
Cartesian grid is shown in Fig.~\ref{c2d500K}, which should be
compared with that for the polar grid shown in the second row of
Fig.~\ref{p2dNvary}, for which the number of particles and softening
length were identical.  Again the curves for separate runs have been
shifted in time so that they all pass through amplitude 0.1 at the
same instant, which is the mean of the set shown.

While there are differences in detail between the two figures, the
mean and spread in the amplitude evolution are quite similar.

\begin{figure}
\includegraphics[width=.57\hsize,angle=270]{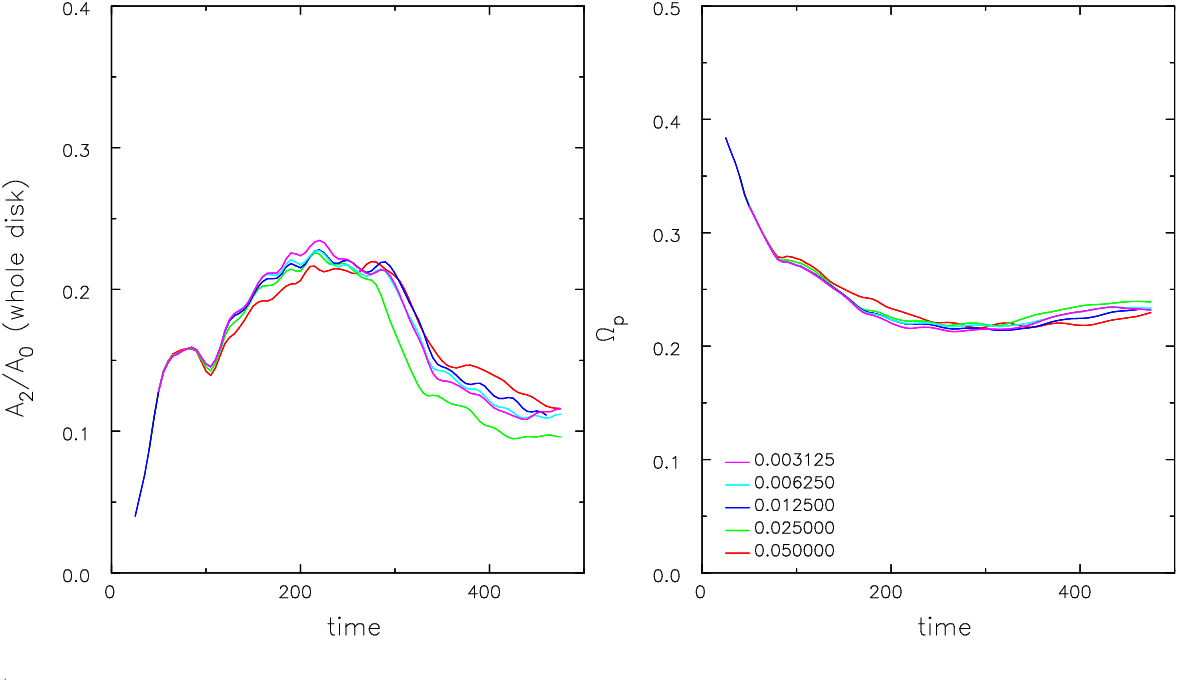}
\includegraphics[width=.57\hsize,angle=270]{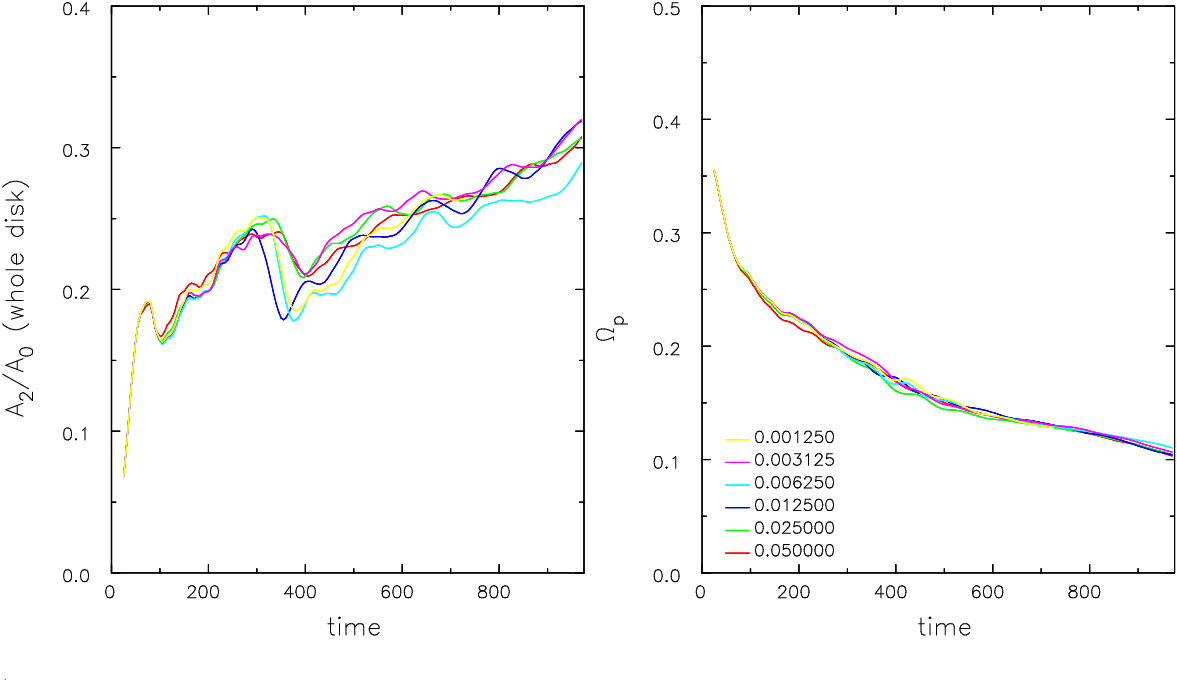}
\caption{Evolution of the bar in two sets of runs with the same random
seeds for the disc particle coordinates.  Numerical parameters are
given in Table~\ref{params} except the time step is varied.  Values
adopted are colour coded as shown.  Upper panels show the evolution
from one set of initial coordinates, lower panels from a second set.}
\label{tstep}
\end{figure}

\subsection{Time Step}
Fig.~\ref{tstep} shows that changing the time step also has little
effect on the evolution.  These tests are for two of the 3D models
shown in Fig.~\ref{basichyb}, one in which the bar slowed at late
times and one in which it did not.  The value of the time step
parameter is varied by a factor of 40 in the case that slowed
strongly.  Small differences in the evolution develop at late times
because the system is chaotic but the deviations do not vary
systematically with the step size.  If the orbital angular frequency
is $\Omega_c$, a particle takes $2\pi/(\Omega_c \Delta t)$ steps for a
circular orbit.  The central value of $\Omega_c \simeq 2$ for our
standard model, implying 250 steps per orbit for the most bound
particles at our standard time-step, and ten times as many for the
shortest step used in Fig.~\ref{tstep}.  In agreement with
\citet{DBS9}, we therefore find no evidence to support the claim by
\citet{KVCQ} that these simulations require $>2000$ time steps per
orbit period for the most tightly bound particles.

We have also verified that the evolution is similarly insensitive to
using a fixed time step for all particles, instead of the more
efficient scheme of employing longer steps at larger distances from
the centre.

\begin{figure}
\includegraphics[width=.57\hsize,angle=270]{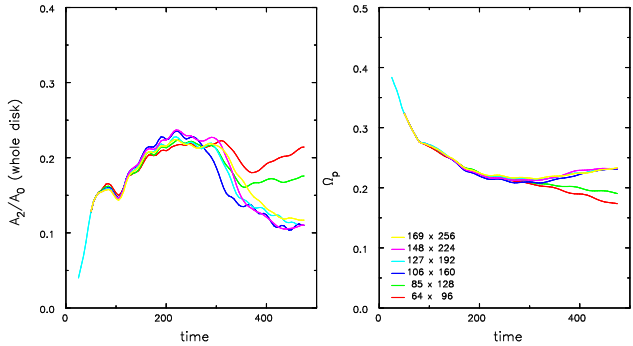}
\caption{Evolution of the bar in 3D runs with the same random seeds
for the disc particle coordinates using different grid sizes.
Numerical parameters are given in Table~\ref{params} except the number
of grid cells used for the 3D polar grid is changed, as indicated by
the line colours.}
\label{grids}
\end{figure}

\subsection{Grid Resolution}
Fig.~\ref{grids} shows the effects of changing the size of the
cylindrical polar grid used for the disc in the hybrid code, keeping
the initial particle coordinates and all other numerical parameters
fixed.  As in other tests, small differences in the evolution develop
at late times but aside from the two coarsest grids, for which the
late time evolution departs systematically from the rest, the results
are quite similar.  We have also found that smaller differences result
when we double or halve the vertical spacing of the grid.  Our
standard grid (Table~\ref{params}) is shown by the cyan line and seems
adequate.

In addition, we checked that the evolution is unaffected (to the same
level of tolerance) by changing the number of active sectoral
harmonics of the polar grid to $m_{\rm max} = 4$ or $m_{\rm max} = 16$
from our standard value of $m_{\rm max} = 8$, or by changing the order
of azimuthal expansion $l_{\rm max}$ and the number of shells $n_r$ of
the spherical polar grid from our standard values of $l_{\rm max}=4$
and $n_r=300$.

\begin{figure}
\includegraphics[width=.57\hsize,angle=270]{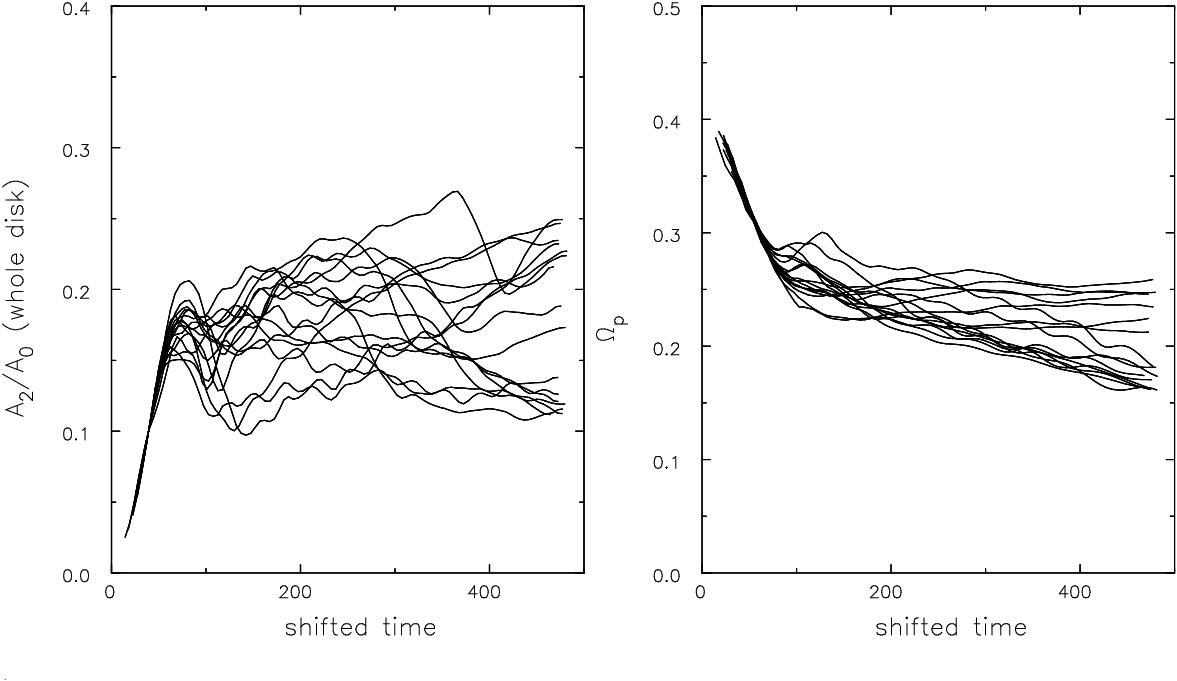}
\includegraphics[width=.57\hsize,angle=270]{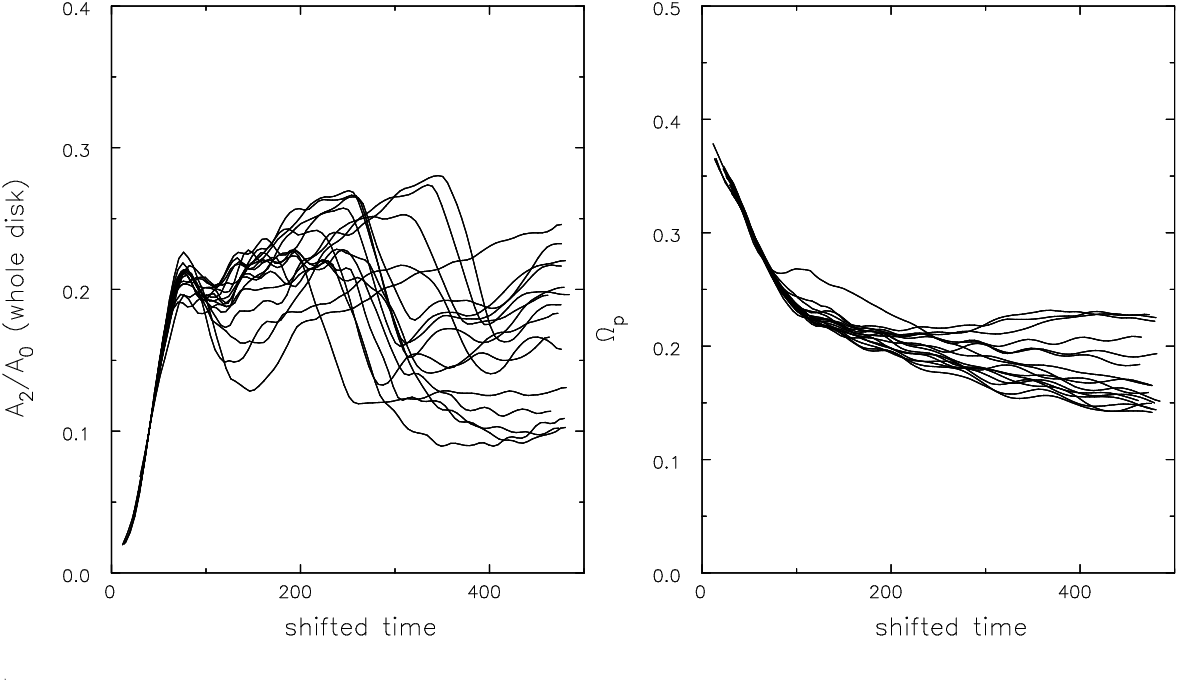}
\caption{Evolution of the bar in 3D runs with the same sets of random
seeds for the disc particles as in Fig.~\ref{basichyb} but using
different softening lengths.  The softening length in the upper and
lower panels is respectively halved and doubled from our standard
value.}
\label{vsoft}
\end{figure}

\subsection{Softening}
Fig.~\ref{vsoft} shows the effects of changing the softening length
for force convolution on the 3D polar grid used for the disc in the
hybrid code, with the initial particle coordinates and other numerical
parameters unchanged from Fig.~\ref{basichyb}.  As reported
elsewhere \citep[][\etc]{Sell81,Sell83,SM94}, the evolution of disc
instabilities is more sensitive to this numerical parameter than
perhaps any other.  The growth rates of both bar-forming modes, and of
bending modes are quite sensitive to the sharpness of short-range
forces.  The effect of a longer softening length (lower panels) is
both to increase the initial peak amplitude of the bar, because the
second mode is more strongly suppressed by softening than is the
dominant, and to make bending instabilities occur later and more
violently.\footnote{The reason is as follows, adapted from
\citet{MS94}: If stars move at speed $u$ in one-dimension over a
ripple of wavenumber $k$, then a condition for a growing bend is that
$ku < \kappa_z$.  Increasing softening reduces $\kappa_z$, which
causes only smaller $k$, or longer wavelength, bends to grow.}  The
effects of a reduction in softening are less systematic, but the extra
virulence of swing-amplified shot noise is the probable cause of more
marked upward fluctuations in the pattern speed evolution and there
are fewer violent buckling events.

Since it is desirable to use the largest value that does not have a
systematic influence on the outcome, these tests show that our
standard value seems a reasonable compromise.

\begin{figure}
\includegraphics[width=.57\hsize,angle=270]{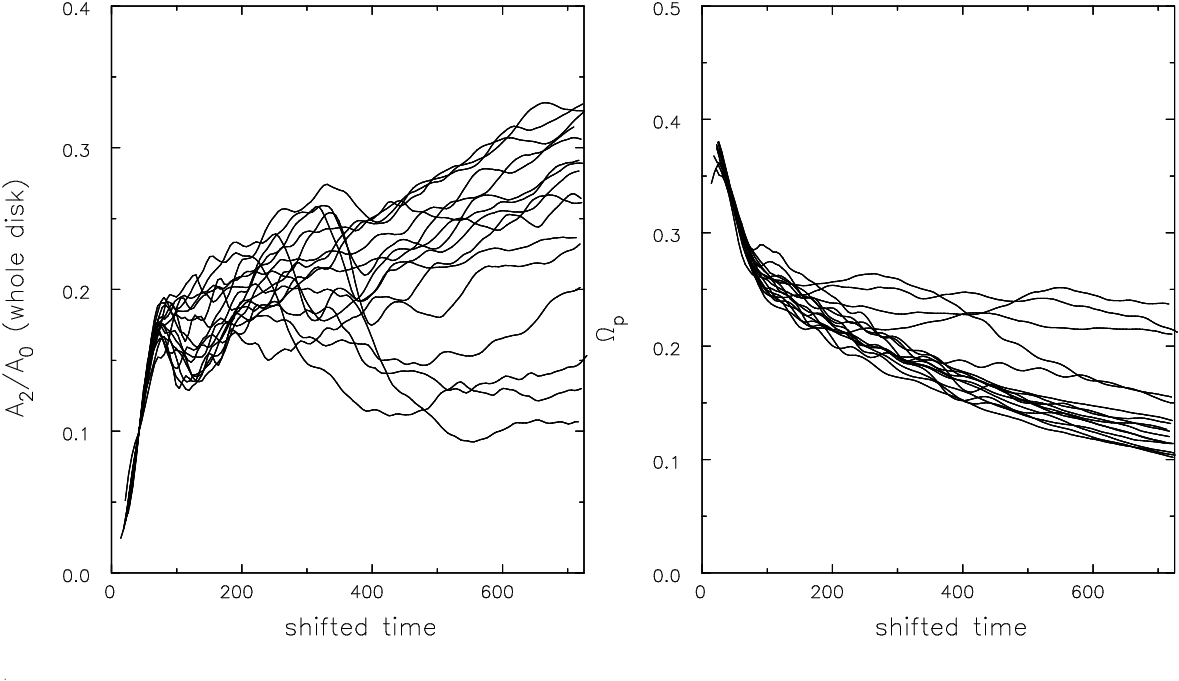}
\includegraphics[width=.57\hsize,angle=270]{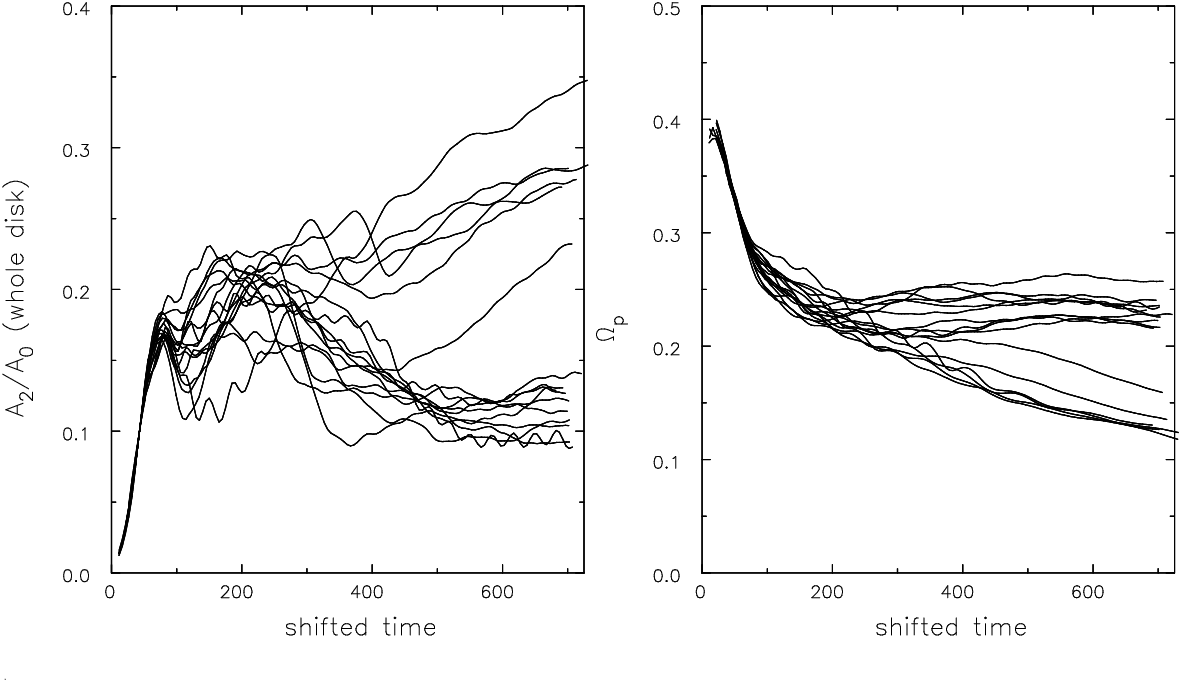}
\caption{Evolution of the bar in runs to test the dependence on the
number of halo particles.  The upper panels used one tenth the number
of unequal mass particles employed in Fig.~\ref{basichyb}, while the
number employed in the lower panels was 10 million.  Other numerical
parameters are held fixed at the values given in Table~\ref{params}.}
\label{vhaloN}
\end{figure}

\subsection{Number of Halo Particles}
Fig.~\ref{vhaloN} shows two sets of runs with different numbers of
unequal mass halo particles, in which the random seeds for the disc
particles were changed.  (We already reported the dependence of the
behaviour on the number of disc particles in Fig.~\ref{p2dNvary}.)
Again, the behaviour in these tests, and in another set with 2.5M equal
mass particles, is qualitatively similar to that shown in
Fig~\ref{basichyb}.  The ranges of final amplitudes and pattern speeds
do not depend on the number of halo particles or whether the masses
are all equal.  There is a trend, in that the fraction of bars that do
not experience strong friction seems to increase with increasing
numbers of halo particles: it is 4/16 for $N_h=2.5 \times 10^5$, 7/16
for $N_h=2.5 \times 10^6$ (Fig.~\ref{basichyb}) and 11/16 for $N_h =
10^7$.  For the experiments with $N_h = 2.5 \times 10^6$ equal mass
particles, the non-slowing fraction is 3/16.

We make use of this trend with the quality of the simulations in the
discussion of Sections~\ref{pselect} \& \ref{discussion}.

\section{Effects of particle selection for the isochrone disc}
Here we illustrate the advantages of careful particle selection for a
simple disc model with well-defined global instabilities.  The value
of a quiet start was already illustrated by comparison of
Figs.~\ref{isoc8dfdetq} \& \ref{isoc8dfdetn} but particles were
deterministically selected from the \DF\ for both sets of simulations.

\begin{figure}
\includegraphics[width=.57\hsize,angle=270]{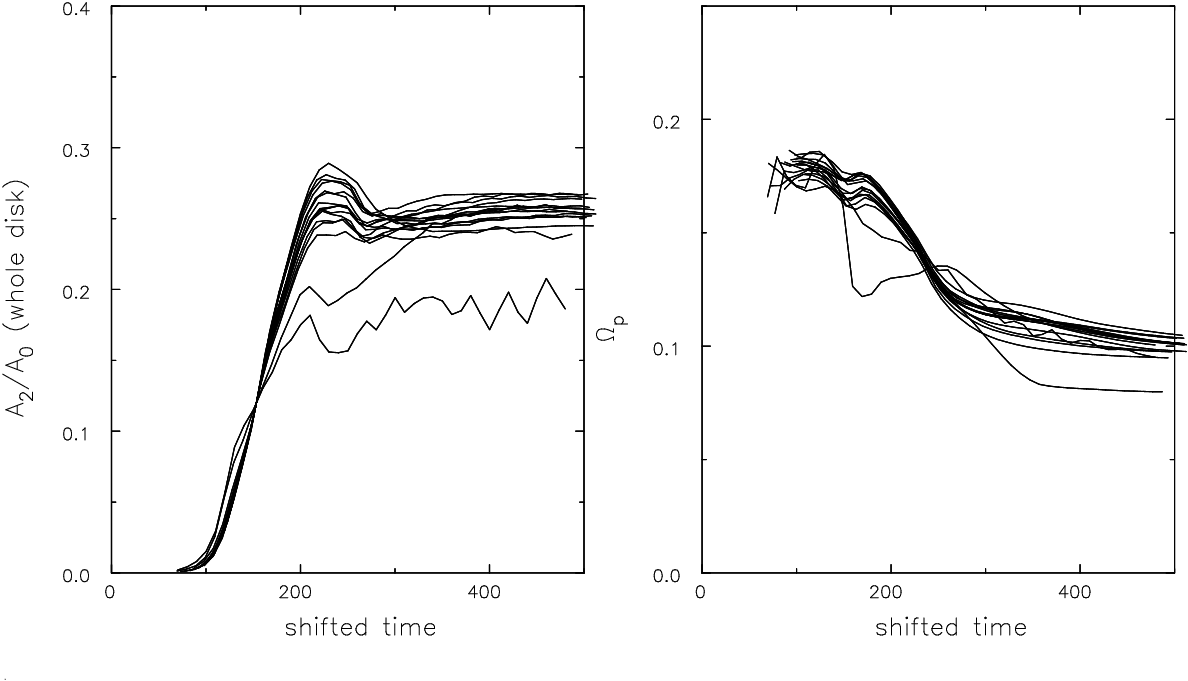}
\caption{Evolution of the bar in the isochrone/8 disc, but instead of
selecting particles deterministically as in Fig.~\ref{isoc8dfdetq}, we
used a simple acceptance/rejection algorithm.  Note the larger spread
in the measured bar properties.}
\label{isoc8dfarjq}
\vspace{0.3cm}
\includegraphics[width=.57\hsize,angle=270]{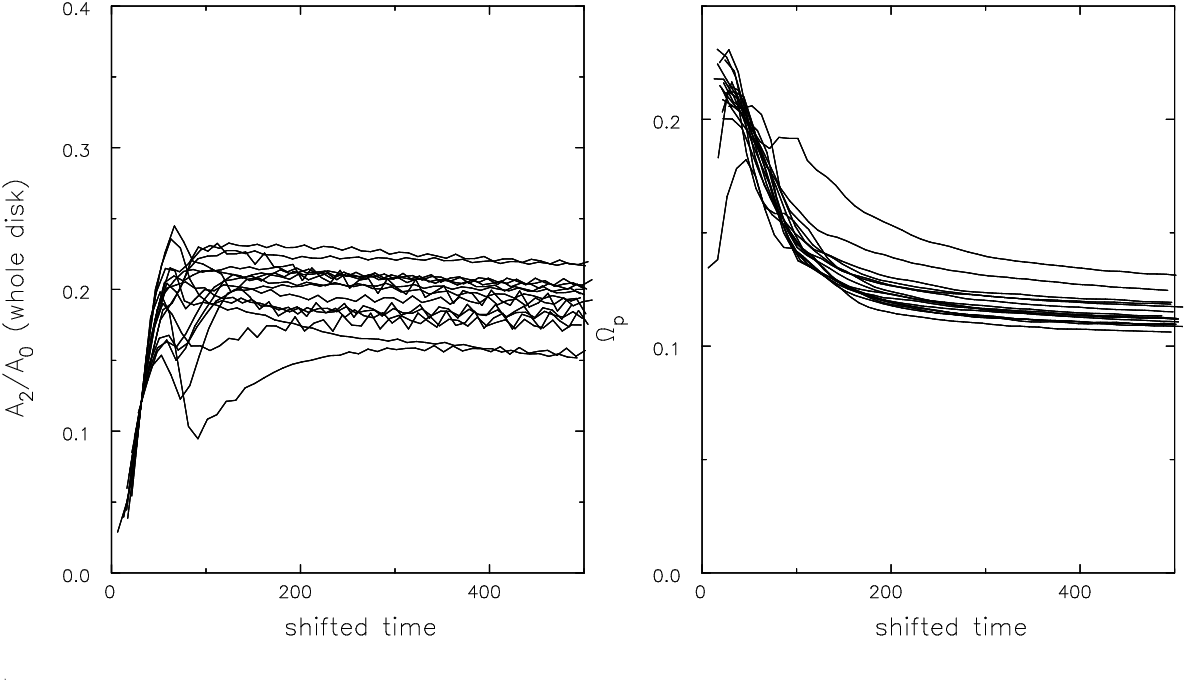}
\caption{Evolution of the bar in a noisy start isochrone disc in which
the non-circular motions were set up crudely rather than selecting
from a \DF.  The value of $Q$ in the initial disc is similar to that
of the initial models in Figs.~\ref{isoc8dfdetq}, \ref{isoc8dfdetn},
\& \ref{isoc8dfarjq}.}
\label{isoc8jsdspn}
\end{figure}

\begin{figure}
\includegraphics[width=.57\hsize,angle=270]{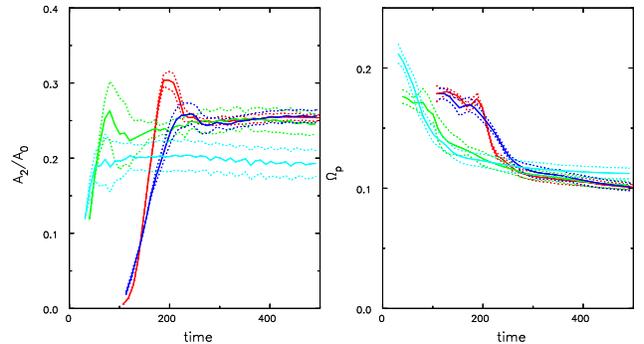}
\caption{Summary of results from in Figs.~\ref{isoc8dfdetq} (red),
\ref{isoc8dfdetn} (green), \ref{isoc8dfarjq} (blue), \&
\ref{isoc8jsdspn} (cyan) in order to illustrate the ranges of
scatter.}
\label{select}
\end{figure}

Fig.~\ref{isoc8dfarjq} shows the consequences of selecting particles
by the commonly-used acceptance/rejection method.  Even though these
experiments still used quiet starts (replicas of each master particle
spaced evenly around a ring), the results are less well behaved: there
is more scatter particularly in the bar amplitude, with one or two
significantly anomalous results.

Fig.~\ref{isoc8jsdspn} shows the results from experiments in which
the set up procedure for the random speeds of the disc particles
stemmed simply from the requirement that $Q=1.2$ everywhere, with the
azimuthal dispersion and asymmetric drift determined by Jeans
equations in the epicycle approximation, as suggested by
\citet{Hern93}.  Although this may be the most commonly used method,
the outcome of such experiments shows the greatest degree of scatter.

The effects of quiet and noisy starts, and other particle selection
issues are summarized in Fig.~\ref{select}.  Generally, experiments
with noisy starts show considerably more scatter than do those with
quiet starts, and deterministic selecting from a \DF\ is superior to
random sampling or not using a \DF\ at all.

\label{lastpage}


\begin{thebibliography}{99}

\def\aap{A\&A}
\def\aj{AJ}
\def\apj{ApJ}
\def\apjl{ApJL}
\def\apjs{ApJS}
\def\araa{ARAA}
\def\jcoph{J. Comp.\ Phys.}
\def\mnras{MNRAS}
\def\nat{Nature}

\bibitem[\protect\citeauthoryear
{Aarseth}{2008}]{Aars08}
Aarseth, S. J. 2008, in ``The Cambridge N-Body Lectures'', Lecture Notes in Physics, {\bf 760} (Springer-Verlag: Berlin Heidelberg), p.~1

\bibitem[\protect\citeauthoryear
{Arfken}{1985}]{Arfk85}
Arfken, G. 1985, {\it Mathematical Methods for Physicists,} 3rd ed.\ (Orlando: Academic Press)

\bibitem[\protect\citeauthoryear
{Athanassoula}{2002}]{Atha02}
Athanassoula, E. 2002, \apjl, {\bf 569}, L83

\bibitem[\protect\citeauthoryear
{Athanassoula}{2003}]{Atha03}
Athanassoula, E. 2003, \mnras, {\bf 341}, 1179

\bibitem[\protect\citeauthoryear
{Beers \etal}{1990}]{Beers}
Beers, T. C., Flynn, K. \& Gebhardt, K. 1990, \aj, {\bf 100}, 32

\bibitem[\protect\citeauthoryear
{Binney \& Tremaine}{2008}]{BT08}
Binney, J. \& Tremaine, S. 2008, {\it Galactic Dynamics\/} 2nd edition (Princeton: Princeton University Press) (BT08)

\bibitem[\protect\citeauthoryear
{Camm}{1950}]{Camm50}
Camm, G. L, 1950, \mnras, {\bf 110}, 305

\bibitem[\protect\citeauthoryear
{Combes \& Sanders}{1981}]{CS81}
Combes, F. \& Sanders, R. H. 1981, \aap, {\bf 96}, 164

\bibitem[\protect\citeauthoryear
{Debattista \& Sellwood}{2000}]{DS00}
Debattista, V. P. \& Sellwood, J. A. 2000, \apj, {\bf 543}, 704

\bibitem[\protect\citeauthoryear
{Dehnen}{2001}]{Dehn01}
Dehnen, W. 2001, \mnras, {\bf 324}, 273

\bibitem[\protect\citeauthoryear
{Diemand \etal}{2004}]{DMS4}
Diemand, J., Moore, B., Stadel, J. 2004, \mnras, {\bf 353}, 624

\bibitem[\protect\citeauthoryear
{Dubinski \etal}{2009}]{DBS9}
Dubinski, J.,  Berentzen, I. \& Shlosman, I. 2009, \apj, {\bf 697}, 293

\bibitem[\protect\citeauthoryear
{Earn \& Sellwood}{1995}]{ES95}
Earn, D. J. D. \& Sellwood, J. A. 1995, \apj, {\bf 451}, 533

\bibitem[\protect\citeauthoryear
{El-Zant \etal}{2004}]{EZ04}
El-Zant, A.,  Hoffman, Y., Primack, J., Combes, F. \& Shlosman, I. 2004, \apjl, {\bf 607}, L75

\bibitem[\protect\citeauthoryear
{Erickson}{1975}]{Eric75}
Erickson, S. A. 1975, PhD thesis, MIT.

\bibitem[\protect\citeauthoryear
{Frenk \etal}{1999}]{Fren99}
Frenk, C. S., \etal\ 1999, \apj, {\bf 525}, 554

\bibitem[\protect\citeauthoryear
{Hernquist}{1990}]{Hern90}
Hernquist, L. 1990, \apj, {\bf 356}, 359

\bibitem[\protect\citeauthoryear
{Hernquist}{1993}]{Hern93}
Hernquist, L. 1993, \apjs, {\bf 86}, 389

\bibitem[\protect\citeauthoryear
{Holley-Bockelmann \etal}{2005}]{HBWK}
Holley-Bockelmann, K., Weinberg, M. \& Katz, N. 2005, \mnras, {\bf 363}, 991

\bibitem[\protect\citeauthoryear
{Jalali}{2007}]{Jala07}
Jalali, M. A. 2007, \apj, {\bf 669}, 218

\bibitem[\protect\citeauthoryear
{Kalnajs}{1976}]{Kaln76}
Kalnajs, A. J. 1976, \apj, {\bf 205}, 751

\bibitem[\protect\citeauthoryear
{Kalnajs}{1978}]{Kaln78}
Kalnajs, A. J. 1978, in {\it Structure and Properties of Nearby Galaxies} IAU Symposium {\bf 77} eds.\ E. M. Berkhuisjen \& R. Wielebinski (Dordrecht:Reidel) p.~113

\bibitem[\protect\citeauthoryear
{Klypin \etal}{2008}]{KVCQ}
Klypin, A., Valenzuela, O, Col\'\i n, P. \& Quinn, T. 2008, arXiv:0808.3422

\bibitem[\protect\citeauthoryear
{Kuijken \& Dubinski}{1995}]{KD95}
Kuijken, K. \& Dubinski, J. 1995, \mnras, {\bf 277}, 1341

\bibitem[\protect\citeauthoryear
{McGlynn}{1984}]{McGl84}
McGlynn, T. A. 1984, \apj, {\bf 281}, 13

\bibitem[\protect\citeauthoryear
{Merritt \& Sellwood}{1994}]{MS94}
Merritt, D. \& Sellwood, J. A. 1994, \apj, {\bf 425}, 551

\bibitem[\protect\citeauthoryear
{Miller}{1964}]{Mill64}
Miller, R. H. 1964, \apj, {\bf 140}, 250

\bibitem[\protect\citeauthoryear
{Monaghan}{1992}]{Mona92}
Monaghan, J. 1992, \araa, {\bf 30}, 543

\bibitem[\protect\citeauthoryear
{Monaghan \& Lattanzio}{1985}]{ML85}
Monaghan, J. \& Lattanzio,  1985, \aap, {\bf 149}, 135

\bibitem[\protect\citeauthoryear
{Power \etal}{2003}]{Powe03}
Power, C., Navarro, J. F., Jenkins, A., Frenk, C. S. \& White, S. D. M. 2003, \mnras, {\bf 338}, 14

\bibitem[\protect\citeauthoryear
{Prendergast \& Tomer}{1970}]{PT70}
Prendergast, K. H. \& Tomer, E. 1970 \aj, {\bf 75}, 674

\bibitem[\protect\citeauthoryear
{Raha \etal}{1991}]{RSJK}
Raha, N., Sellwood, J. A., James, R. A. \& Kahn, F. D. 1991, \nat, {\bf 352}, 411

\bibitem[\protect\citeauthoryear
{Romeo}{1992}]{Rome92}
Romeo, A. 1992, \mnras, {\bf 256}, 307

\bibitem[\protect\citeauthoryear
{Sellwood}{1981}]{Sell81}
Sellwood, J. A. 1981, \aap, {\bf 99}, 362

\bibitem[\protect\citeauthoryear
{Sellwood}{1983}]{Sell83}
Sellwood, J. A. 1983, J. Comp.\ Phys., {\bf 50}, 337

\bibitem[\protect\citeauthoryear
{Sellwood}{1985}]{Sell85}
Sellwood, J. A. 1985, \mnras, {\bf 217}, 127

\bibitem[\protect\citeauthoryear
{Sellwood}{1989}]{Sell89}
Sellwood, J. A. 1989, \mnras, {\bf 238}, 115

\bibitem[\protect\citeauthoryear
{Sellwood}{2003}]{Sell03}
Sellwood, J. A. 2003, \apj, {\bf 587}, 638

\bibitem[\protect\citeauthoryear
{Sellwood}{2008}]{Sell08}
Sellwood, J. A. 2008, \apj, {\bf 679}, 379

\bibitem[\protect\citeauthoryear
{Sellwood \& Athanassoula}{1986}]{SA86}
Sellwood, J. A. \& Athanassoula, E. 1986, \mnras, {\bf 221}, 195

\bibitem[\protect\citeauthoryear
{Sellwood \& Debattista}{2006}]{SD06}
Sellwood, J. A. \& Debattista, V. P. 2006, \apj, {\bf 639}, 868

\bibitem[\protect\citeauthoryear
{Sellwood \& Evans}{2001}]{SE01}
Sellwood, J. A. \& Evans, N. W. 2001, \apj, {\bf 546}, 176

\bibitem[\protect\citeauthoryear
{Sellwood \& McGaugh}{2005}]{SM05}
Sellwood, J. A. \& McGaugh, S. S. 2005, \apj, {\bf 634}, 70

\bibitem[\protect\citeauthoryear
{Sellwood \& Merritt}{1994}]{SM94}
Sellwood, J. A. \& Merritt, D. 1994, \apj, {\bf 425}, 530

\bibitem[\protect\citeauthoryear
{Shlosman \& Noguchi}{1993}]{SN93}
Shlosman, I. \& Noguchi, M. 1993, \apj, {\bf 414}, 474

\bibitem[\protect\citeauthoryear
{Shu}{1969}]{Shu69}
Shu, F. H. 1969, \apj, {\bf 158}, 505

\bibitem[\protect\citeauthoryear
{Spitzer}{1942}]{Spit42}
Spitzer, L. 1942, \apj, {\bf 95}, 329

\bibitem[\protect\citeauthoryear
{Stadel}{2001}]{Stad01}
Stadel, J. G., 2001, Ph.D. thesis, University of Washington.

\bibitem[\protect\citeauthoryear
{Toomre}{1964}]{Toom64}
Toomre, A. 1964, \apj, {\bf 139}, 1217

\bibitem[\protect\citeauthoryear
{Toomre}{1981}]{Toom81}
Toomre, A. 1981, in {\it The Structure and Evolution of Normal Galaxies}, eds.\ S. M. Fall \& D. Lynden-Bell (Cambridge: Cambridge University Press), p.~111

\bibitem[\protect\citeauthoryear
{Valenzuela \& Klypin}{2003}]{VK03}
Valenzuela, O. \& Klypin, A. 2003, \mnras, {\bf 345}, 406

\bibitem[\protect\citeauthoryear
{Vandervoort}{1970}]{Vand70}
Vandervoort, P. O. 1970, \apj, {\bf 161}, 87

\bibitem[\protect\citeauthoryear
{Weinberg \& Katz}{2007}]{WK07}
Weinberg, M. D. \& Katz, N. 2007, \mnras, {\bf 375}, 460

\bibitem[\protect\citeauthoryear
{Young}{1980}]{Youn80}
Young, P. 1980, \apj, {\bf 242}, 1232

\bibitem[\protect\citeauthoryear
{Zhang \& Magorrian}{2008}]{ZM08}
Zhang, M. \& Magorrian, J, 2008, \mnras, {\bf 387}, 1719

\end{thebibliography}
\end{document}